\documentclass[sigconf]{acmart}
\acmConference[EASE 2025]{The 29th International Conference on Evaluation and Assessment in Software Engineering}{17–20 June, 2025}{Istanbul, Türkiye}

\usepackage{listings}
\lstdefinestyle{interfaces}{
  float=t,
  floatplacement=t,
  abovecaptionskip=-5pt
}

\usepackage{microtype}
\usepackage[utf8]{inputenc}
\usepackage[capitalise,noabbrev]{cleveref}
\usepackage{tabularx}
\usepackage{subcaption}
\usepackage{enumitem}
\usepackage{booktabs}
\usepackage{xcolor}
\usepackage{tcolorbox}
\usepackage{tikz}

\usepackage{xspace}
\usepackage{multirow}
\usepackage{afterpage}

\usepackage{graphicx}
\usepackage{balance}
\usepackage{float}

\begin{document}
\definecolor{mGreen}{rgb}{0,0.6,0}
\definecolor{mGray}{rgb}{0.5,0.5,0.5}
\definecolor{mPurple}{rgb}{0.58,0,0.82}
\definecolor{backgroundColour}{rgb}{0.99,0.99,0.97}

\definecolor{backgroundColourBlue}{rgb}{0.61, 0.87, 1.0} 
\definecolor{backgroundColourGreen}{rgb}{0.4, 1.0, 0.0} 
\definecolor{backgroundColourYellow}{rgb}{1.0, 1.0, 0.2} 

\lstdefinestyle{CStyle}{
    commentstyle=\color{mGreen},
    keywordstyle=\color{magenta},
    numberstyle=\color{mGray},
    stringstyle=\color{mPurple},
    basicstyle=\tt\fontsize{8}{9}\selectfont,
    breakatwhitespace=false,         
    breaklines=true,                 
    captionpos=b,
		keepspaces=true,                 
    numbers=none,
    stepnumber=1,
		firstnumber=1,
    numbersep=3pt,                  
    showspaces=false,                
    showstringspaces=false,
    showtabs=false,                  
    tabsize=2,
    language=C
}

\lstdefinestyle{CStyleMM}{
    commentstyle=\color{mGreen},
    keywordstyle=\color{magenta},
    numberstyle=\color{mGray},
    stringstyle=\color{mPurple},
    basicstyle=\tt\fontsize{6}{7}\selectfont,
    breakatwhitespace=false,         
    breaklines=true,                 
    captionpos=b,
		keepspaces=true,                 
    numbers=left,
    stepnumber=1,
		firstnumber=1,
    numbersep=3pt,                  
    showspaces=false,                
    showstringspaces=false,
    showtabs=false,                  
    tabsize=2,
    language=C
}

\lstdefinestyle{CStyle1A}{
    commentstyle=\color{mGreen},
    keywordstyle=\color{magenta},
    numberstyle=\color{mGray},
    stringstyle=\color{mPurple},
    basicstyle=\tt\fontsize{8}{9}\selectfont,
    breakatwhitespace=false,         
    breaklines=true,                 
    captionpos=b,
		keepspaces=true,                 
    numbers=left,
    stepnumber=1,
		firstnumber=1,
    numbersep=3pt,                  
    showspaces=false,                
    showstringspaces=false,
    showtabs=false,                  
    belowskip=-4pt,
     tabsize=2,
    language=C
}

\lstdefinestyle{CStyleA}{
    commentstyle=\color{mGreen},
    keywordstyle=\color{magenta},
    numberstyle=\color{mGray},
    stringstyle=\color{mPurple},
    basicstyle=\tt\fontsize{8}{9}\selectfont,
    breakatwhitespace=false,         
    breaklines=true,                 
    captionpos=b,
		keepspaces=true,                 
    numbers=left,
    stepnumber=1,
		firstnumber=4,
    numbersep=3pt,                  
    showspaces=false,                
    showstringspaces=false,
    showtabs=false,                  
    tabsize=2,
		aboveskip=-2pt,
    belowskip=-4pt,
     language=C
}

\lstdefinestyle{CStyleA2}{
    backgroundcolor=\color{backgroundColour},   
    commentstyle=\color{mGreen},
    keywordstyle=\color{magenta},
    numberstyle=\color{mGray},
    stringstyle=\color{mPurple},
    basicstyle=\tt\fontsize{8}{9}\selectfont,
    breakatwhitespace=false,         
    breaklines=true,                 
    captionpos=b,
		keepspaces=true,                 
    numbers=left,
    stepnumber=1,
		firstnumber=37,
    numbersep=3pt,                  
    showspaces=false,                
    showstringspaces=false,
    showtabs=false,                  
    tabsize=2,
		aboveskip=-2pt,
    belowskip=-4pt,
     language=C
}

\lstdefinestyle{CStyleA3}{
    backgroundcolor=\color{backgroundColour},   
    commentstyle=\color{mGreen},
    keywordstyle=\color{magenta},
    numberstyle=\color{mGray},
    stringstyle=\color{mPurple},
    basicstyle=\tt\fontsize{8}{9}\selectfont,
    breakatwhitespace=false,         
    breaklines=true,                 
    captionpos=b,
		keepspaces=true,                 
    numbers=left,
    stepnumber=1,
		firstnumber=44,
    numbersep=3pt,                  
    showspaces=false,                
    showstringspaces=false,
    showtabs=false,                  
    tabsize=2,
		aboveskip=-2pt,
    belowskip=-4pt,
     language=C
}

\lstdefinestyle{CStyleA4}{
    backgroundcolor=\color{backgroundColour},   
    commentstyle=\color{mGreen},
    keywordstyle=\color{magenta},
    numberstyle=\color{mGray},
    stringstyle=\color{mPurple},
    basicstyle=\tt\fontsize{8}{9}\selectfont,
    breakatwhitespace=false,         
    breaklines=true,                 
    captionpos=b,
		keepspaces=true,                 
    numbers=left,
    stepnumber=1,
		firstnumber=52,
    numbersep=3pt,                  
    showspaces=false,                
    showstringspaces=false,
    showtabs=false,                  
    tabsize=2,
		aboveskip=-2pt,
    belowskip=-4pt,
     language=C
}
\lstdefinestyle{CStyleA5}{
    backgroundcolor=\color{backgroundColour},   
    commentstyle=\color{mGreen},
    keywordstyle=\color{magenta},
    numberstyle=\color{mGray},
    stringstyle=\color{mPurple},
    basicstyle=\tt\fontsize{8}{9}\selectfont,
    breakatwhitespace=false,         
    breaklines=true,                 
    captionpos=b,
		keepspaces=true,                 
    numbers=left,
    stepnumber=1,
		firstnumber=58,
    numbersep=3pt,                  
    showspaces=false,                
    showstringspaces=false,
    showtabs=false,                  
    tabsize=2,
		aboveskip=-2pt,
    belowskip=-4pt,
     language=C
}
\lstdefinestyle{CStyle1}{
    backgroundcolor=\color{backgroundColourBlue},   
    commentstyle=\color{mGreen},
    keywordstyle=\color{magenta},
    numberstyle=\color{mGray},
    stringstyle=\color{mPurple},
    basicstyle=\tt\fontsize{8}{9}\selectfont,
    breakatwhitespace=false,         
    breaklines=true,                 
    captionpos=b,
		keepspaces=true,                 
    numbers=left,
    stepnumber=1,
		firstnumber=2,
    numbersep=3pt,                  
    showspaces=false,                
    showstringspaces=false,
    showtabs=false,                  
    tabsize=2,
		aboveskip=-2pt,
    belowskip=-4pt,
    language=C
}


\lstdefinestyle{CStyleB}{
    backgroundcolor=\color{backgroundColourBlue},   
    commentstyle=\color{mGreen},
    keywordstyle=\color{magenta},
    numberstyle=\color{mGray},
    stringstyle=\color{mPurple},
    basicstyle=\tt\fontsize{8}{9}\selectfont,
    breakatwhitespace=false,         
    captionpos=b,
		keepspaces=true,                 
    numbers=left,
    stepnumber=1,
		firstnumber=3,
    numbersep=3pt,                  
    showspaces=false,                
    showstringspaces=false,
    showtabs=false,                  
    tabsize=2,
    language=C
}

\lstdefinestyle{CStyleB2}{
    backgroundcolor=\color{backgroundColourBlue},   
    commentstyle=\color{mGreen},
    keywordstyle=\color{magenta},
    numberstyle=\color{mGray},
    stringstyle=\color{mPurple},
    basicstyle=\tt\fontsize{8}{9}\selectfont,
    breakatwhitespace=false,         
    breaklines=true,                 
    captionpos=b,
		keepspaces=true,                 
    numbers=left,
    stepnumber=1,
		firstnumber=22,
    numbersep=3pt,                  
    showspaces=false,                
    showstringspaces=false,
    showtabs=false,                  
    tabsize=2,
    language=C
}

\lstdefinestyle{CStyleB3}{
    backgroundcolor=\color{backgroundColourBlue},   
    commentstyle=\color{mGreen},
    keywordstyle=\color{magenta},
    numberstyle=\color{mGray},
    stringstyle=\color{mPurple},
    basicstyle=\tt\fontsize{8}{9}\selectfont,
    breakatwhitespace=false,         
    breaklines=true,                 
    captionpos=b,
		keepspaces=true,                 
    numbers=left,
    stepnumber=1,
		firstnumber=40,
    numbersep=3pt,                  
    showspaces=false,                
    showstringspaces=false,
    showtabs=false,                  
    tabsize=2,
    language=C
}

\lstdefinestyle{CStyleB3-A}{
    backgroundcolor=\color{backgroundColourBlue},   
    commentstyle=\color{mGreen},
    keywordstyle=\color{magenta},
    numberstyle=\color{mGray},
    stringstyle=\color{mPurple},
    basicstyle=\tt\fontsize{8}{9}\selectfont,
    breakatwhitespace=false,         
    breaklines=true,                 
    captionpos=b,
		keepspaces=true,                 
    numbers=left,
    stepnumber=1,
		firstnumber=43,
    numbersep=3pt,                  
    showspaces=false,                
    showstringspaces=false,
    showtabs=false,                  
    tabsize=2,
    language=C
}

\lstdefinestyle{CStyleB4}{
    backgroundcolor=\color{backgroundColourBlue},   
    commentstyle=\color{mGreen},
    keywordstyle=\color{magenta},
    numberstyle=\color{mGray},
    stringstyle=\color{mPurple},
    basicstyle=\tt\fontsize{8}{9}\selectfont,
    breakatwhitespace=false,         
    breaklines=true,                 
    captionpos=b,
		keepspaces=true,                 
    numbers=left,
    stepnumber=1,
		firstnumber=47,
    numbersep=3pt,                  
    showspaces=false,                
    showstringspaces=false,
    showtabs=false,                  
    tabsize=2,
    language=C
}

\lstdefinestyle{CStyleB4-A}{
    backgroundcolor=\color{backgroundColourBlue},   
    commentstyle=\color{mGreen},
    keywordstyle=\color{magenta},
    numberstyle=\color{mGray},
    stringstyle=\color{mPurple},
    basicstyle=\tt\fontsize{8}{9}\selectfont,
    breakatwhitespace=false,         
    breaklines=true,                 
    captionpos=b,
		keepspaces=true,                 
    numbers=left,
    stepnumber=1,
		firstnumber=51,
    numbersep=3pt,                  
    showspaces=false,                
    showstringspaces=false,
    showtabs=false,                  
    tabsize=2,
    language=C
}\lstdefinestyle{CStyleB5}{
    backgroundcolor=\color{backgroundColourBlue},   
    commentstyle=\color{mGreen},
    keywordstyle=\color{magenta},
    numberstyle=\color{mGray},
    stringstyle=\color{mPurple},
    basicstyle=\tt\fontsize{8}{9}\selectfont,
    breakatwhitespace=false,         
    breaklines=true,                 
    captionpos=b,
		keepspaces=true,                 
    numbers=left,
    stepnumber=1,
		firstnumber=55,
    numbersep=3pt,                  
    showspaces=false,                
    showstringspaces=false,
    showtabs=false,                  
    tabsize=2,
    language=C
}

\lstdefinestyle{CStyleC1}{
    backgroundcolor=\color{backgroundColourGreen},   
    commentstyle=\color{mGreen},
    keywordstyle=\color{magenta},
    numberstyle=\color{mGray},
    stringstyle=\color{mPurple},
    basicstyle=\tt\fontsize{8}{9}\selectfont,
    breakatwhitespace=false,         
    captionpos=b,
		keepspaces=true,                 
    numbers=left,
    stepnumber=1,
		firstnumber=42,
    numbersep=3pt,                  
    showspaces=false,                
    showstringspaces=false,
    showtabs=false,                  
    tabsize=2,
    language=C,
		aboveskip=-4pt,
    belowskip=-4pt
 }

\lstdefinestyle{CStyleC2}{
    backgroundcolor=\color{backgroundColourGreen},   
    commentstyle=\color{mGreen},
    keywordstyle=\color{magenta},
    numberstyle=\color{mGray},
		aboveskip=-4pt,
    belowskip=-4pt,
    stringstyle=\color{mPurple},
    basicstyle=\tt\fontsize{8}{9}\selectfont,
    breakatwhitespace=false,         
    breaklines=true,                 
    captionpos=b,
		keepspaces=true,                 
    numbers=left,
    stepnumber=1,
		firstnumber=49,
    numbersep=3pt,                  
    showspaces=false,                
    showstringspaces=false,
    showtabs=false,                  
    tabsize=2,
    language=C
}
\lstdefinestyle{CStyleC3}{
    backgroundcolor=\color{backgroundColourGreen},   
    commentstyle=\color{mGreen},
    keywordstyle=\color{magenta},
    numberstyle=\color{mGray},
    stringstyle=\color{mPurple},
    basicstyle=\tt\fontsize{8}{9}\selectfont,
    breakatwhitespace=false,         
		aboveskip=-4pt,
    breaklines=true,                 
    captionpos=b,
		keepspaces=true,                 
    numbers=left,
    stepnumber=1,
		firstnumber=57,
    numbersep=3pt,                  
    showspaces=false,                
    showstringspaces=false,
    showtabs=false,                  
    tabsize=2,
    language=C
}

\lstdefinestyle{CStyle3}{
    backgroundcolor=\color{backgroundColourGreen},   
    commentstyle=\color{mGreen},
    keywordstyle=\color{magenta},
    numberstyle=\color{mGray},
    stringstyle=\color{mPurple},
    basicstyle=\tt\fontsize{8}{9}\selectfont,
    breakatwhitespace=false,         
    breaklines=false,                 
    captionpos=b,
		keepspaces=true,                 
    numbers=left,
    stepnumber=1,
		firstnumber=2,
    numbersep=3pt,                  
    showspaces=false,                
    showstringspaces=false,
    showtabs=false,                  
    tabsize=2,
		aboveskip=-2pt,
    language=C
}
\lstdefinestyle{CStyle31}{
    backgroundcolor=\color{backgroundColourGreen},   
    commentstyle=\color{mGreen},
    keywordstyle=\color{magenta},
    numberstyle=\color{mGray},
    stringstyle=\color{mPurple},
    basicstyle=\tt\fontsize{6.5}{7.5}\selectfont,
    breakatwhitespace=false,         
    breaklines=false,                 
    captionpos=b,
		keepspaces=true,                 
    numbers=left,
    stepnumber=1,
		firstnumber=5,
    numbersep=3pt,                  
    showspaces=false,                
    showstringspaces=false,
    showtabs=false,                  
    tabsize=2,
		aboveskip=-2pt,
    language=C
}

\lstdefinestyle{CStyle32}{
    backgroundcolor=\color{backgroundColourGreen},   
    commentstyle=\color{mGreen},
    keywordstyle=\color{magenta},
    numberstyle=\color{mGray},
    stringstyle=\color{mPurple},
    basicstyle=\tt\fontsize{6.5}{7.5}\selectfont,
    breakatwhitespace=false,         
    breaklines=false,                 
    captionpos=b,
		keepspaces=true,                 
    numbers=left,
    stepnumber=1,
		firstnumber=17,
    numbersep=3pt,                  
    showspaces=false,                
    showstringspaces=false,
    showtabs=false,                  
    tabsize=2,
		aboveskip=-2pt,
    language=C
}

\lstdefinestyle{CStyle33}{
    backgroundcolor=\color{backgroundColourGreen},   
    commentstyle=\color{mGreen},
    keywordstyle=\color{magenta},
    numberstyle=\color{mGray},
    stringstyle=\color{mPurple},
    basicstyle=\tt\fontsize{6.5}{7.5}\selectfont,
    breakatwhitespace=false,         
    breaklines=false,                 
    captionpos=b,
		keepspaces=true,                 
    numbers=left,
    stepnumber=1,
		firstnumber=22,
    numbersep=3pt,                  
    showspaces=false,                
    showstringspaces=false,
    showtabs=false,                  
    tabsize=2,
		aboveskip=-2pt,
    language=C
}

\lstdefinestyle{CStyle13}{
    backgroundcolor=\color{backgroundColour},   
    commentstyle=\color{mGreen},
    keywordstyle=\color{magenta},
    numberstyle=\color{mGray},
    stringstyle=\color{mPurple},
    basicstyle=\tt\fontsize{7}{8}\selectfont,
    breakatwhitespace=false,         
    breaklines=true,                 
    captionpos=b,
		keepspaces=true,                 
    numbers=left,
    stepnumber=1,
		firstnumber=4,
    numbersep=3pt,                  
    showspaces=false,                
    showstringspaces=false,
    showtabs=false,                  
    tabsize=2,
    language=C
}

\lstdefinestyle{CStyle14}{
    backgroundcolor=\color{backgroundColour},   
    commentstyle=\color{mGreen},
    keywordstyle=\color{magenta},
    numberstyle=\color{mGray},
    stringstyle=\color{mPurple},
    basicstyle=\tt\fontsize{7}{8}\selectfont,
    breakatwhitespace=false,         
    breaklines=true,                 
    captionpos=b,
		keepspaces=true,                 
    numbers=left,
    stepnumber=1,
		firstnumber=12,
    numbersep=3pt,                  
    showspaces=false,                
    showstringspaces=false,
    showtabs=false,                  
    tabsize=2,
    language=C
}

\lstdefinestyle{CStyle15}{
    backgroundcolor=\color{backgroundColourBlue},   
    commentstyle=\color{mGreen},
    keywordstyle=\color{magenta},
    numberstyle=\color{mGray},
    stringstyle=\color{mPurple},
    basicstyle=\tt\fontsize{6.5}{7.5}\selectfont,
    breakatwhitespace=false,         
    breaklines=true,                 
    captionpos=b,
		keepspaces=true,                 
    numbers=left,
    stepnumber=1,
		firstnumber=16,
    numbersep=3pt,                  
    showspaces=false,                
    showstringspaces=false,
    showtabs=false,                  
    tabsize=2,
		aboveskip=-2pt,
    language=C
}

\lstdefinestyle{CStyle16}{
    backgroundcolor=\color{backgroundColourBlue},   
    commentstyle=\color{mGreen},
    keywordstyle=\color{magenta},
    numberstyle=\color{mGray},
    stringstyle=\color{mPurple},
    basicstyle=\tt\fontsize{6.5}{7.5}\selectfont,
    breakatwhitespace=false,         
    breaklines=true,                 
    captionpos=b,
		keepspaces=true,                 
    numbers=left,
    stepnumber=1,
		firstnumber=18,
    numbersep=3pt,                  
    showspaces=false,                
    showstringspaces=false,
    showtabs=false,                  
    tabsize=2,
		aboveskip=-2pt,
    language=C
}
\lstdefinestyle{CStyle12}{
    backgroundcolor=\color{backgroundColour},   
    commentstyle=\color{mGreen},
    keywordstyle=\color{magenta},
    numberstyle=\color{mGray},
    stringstyle=\color{mPurple},
    basicstyle=\tt\fontsize{7}{8}\selectfont,
    breakatwhitespace=false,         
    breaklines=true,                 
    captionpos=b,
		keepspaces=true,                 
    numbers=left,
    stepnumber=1,
		firstnumber=16,
    numbersep=3pt,                  
    showspaces=false,                
    showstringspaces=false,
    showtabs=false,                  
    tabsize=2,
    language=C
}
\lstdefinestyle{CStyle2}{
    backgroundcolor=\color{backgroundColour},   
    commentstyle=\color{mGreen},
    keywordstyle=\color{magenta},
    numberstyle=\color{mGray},
    stringstyle=\color{mPurple},
    basicstyle=\tt\fontsize{6.5}{7.5}\selectfont,
    breakatwhitespace=false,         
    breaklines=true,                 
    captionpos=b,
		keepspaces=true,                 
    numbers=left,
    stepnumber=1,
		firstnumber=10,
    numbersep=3pt,                  
    showspaces=false,                
    showstringspaces=false,
    showtabs=false,                  
    tabsize=2,
    language=C
}

\lstdefinestyle{CStyle4}{
    backgroundcolor=\color{backgroundColour},   
    commentstyle=\color{mGreen},
    keywordstyle=\color{magenta},
    numberstyle=\color{mGray},
    stringstyle=\color{mPurple},
    basicstyle=\tt\fontsize{6.5}{7.5}\selectfont,
    breakatwhitespace=false,         
    breaklines=true,                 
    captionpos=b,
		keepspaces=true,                 
    numbers=left,
    stepnumber=1,
		firstnumber=12,
    numbersep=3pt,                  
    showspaces=false,                
    showstringspaces=false,
    showtabs=false,                  
    tabsize=2,
		aboveskip=-3pt,
    belowskip=-3pt,
    language=C
}

\lstdefinestyle{CStyle5}{
    backgroundcolor=\color{backgroundColourGreen},   
    commentstyle=\color{mGreen},
    keywordstyle=\color{magenta},
    numberstyle=\color{mGray},
    stringstyle=\color{mPurple},
    basicstyle=\tt\fontsize{6.5}{7.5}\selectfont,
    breakatwhitespace=false,         
    breaklines=true,                 
    captionpos=b,
		keepspaces=true,                 
    numbers=left,
    stepnumber=1,
		firstnumber=10,
    numbersep=3pt,                  
    showspaces=false,                
    showstringspaces=false,
    showtabs=false,                  
    tabsize=2,
		aboveskip=-1.5pt,
    belowskip=-3pt,
    language=C
}
\lstdefinestyle{CStyle7}{
    backgroundcolor=\color{backgroundColour},   
    commentstyle=\color{mGreen},
    keywordstyle=\color{magenta},
    numberstyle=\color{mGray},
    stringstyle=\color{mPurple},
    basicstyle=\tt\fontsize{6.5}{7.5}\selectfont,
    breakatwhitespace=false,         
    breaklines=true,                 
    captionpos=b,
		keepspaces=true,                 
    numbers=left,
    stepnumber=1,
		firstnumber=12,
    numbersep=3pt,                  
    showspaces=false,                
    showstringspaces=false,
    showtabs=false,                  
    tabsize=2,
    language=C
}

\lstdefinestyle{CStyle2A1}{
    backgroundcolor=\color{backgroundColourBlue},   
    commentstyle=\color{mGreen},
    keywordstyle=\color{magenta},
    numberstyle=\color{mGray},
    stringstyle=\color{mPurple},
    basicstyle=\tt\fontsize{6.5}{7.5}\selectfont,
    breakatwhitespace=false,         
    breaklines=true,                 
    captionpos=b,
		keepspaces=true,                 
    numbers=left,
    stepnumber=1,
		firstnumber=2,
    numbersep=3pt,                  
    showspaces=false,                
    showstringspaces=false,
    showtabs=false,                  
    tabsize=2,
		aboveskip=-8pt,
    belowskip=-3pt,
    language=C
}

\lstdefinestyle{CStyle2A2}{
    backgroundcolor=\color{backgroundColour},   
    commentstyle=\color{mGreen},
    keywordstyle=\color{magenta},
    numberstyle=\color{mGray},
    stringstyle=\color{mPurple},
    basicstyle=\tt\fontsize{6.5}{7.5}\selectfont,
    breakatwhitespace=false,         
    breaklines=true,                 
    captionpos=b,
		keepspaces=true,                 
    numbers=left,
    stepnumber=1,
		firstnumber=4,
    numbersep=3pt,                  
    showspaces=false,                
    showstringspaces=false,
    showtabs=false,                  
    tabsize=2,
    language=C
}

\lstdefinestyle{CStyle2A3}{
    backgroundcolor=\color{backgroundColourGreen},   
    commentstyle=\color{mGreen},
    keywordstyle=\color{magenta},
    numberstyle=\color{mGray},
    stringstyle=\color{mPurple},
    basicstyle=\tt\fontsize{6.5}{7.5}\selectfont,
    breakatwhitespace=false,         
    breaklines=true,                 
    captionpos=b,
		keepspaces=true,                 
    numbers=left,
    stepnumber=1,
		firstnumber=3,
    numbersep=3pt,                  
    showspaces=false,                
    showstringspaces=false,
    showtabs=false,                  
    tabsize=2,
	  aboveskip=-8pt,
    belowskip=-3pt,
    language=C
}

\lstdefinestyle{CStyle2A4}{
    backgroundcolor=\color{backgroundColour},   
    commentstyle=\color{mGreen},
    keywordstyle=\color{magenta},
    numberstyle=\color{mGray},
    stringstyle=\color{mPurple},
    basicstyle=\tt\fontsize{6.5}{7.5}\selectfont,
    breakatwhitespace=false,         
    breaklines=true,                 
    captionpos=b,
		keepspaces=true,                 
    numbers=left,
    stepnumber=1,
		firstnumber=4,
    numbersep=3pt,                  
    showspaces=false,                
    showstringspaces=false,
    showtabs=false,                  
    tabsize=2,
    language=C
}


\lstdefinestyle{CStyle2B1}{
    backgroundcolor=\color{backgroundColourGreen},   
    commentstyle=\color{mGreen},
    keywordstyle=\color{magenta},
    numberstyle=\color{mGray},
    stringstyle=\color{mPurple},
    basicstyle=\tt\fontsize{6.5}{7.5}\selectfont,
    breakatwhitespace=false,         
    breaklines=true,                 
    captionpos=b,
		keepspaces=true,                 
    numbers=left,
    stepnumber=1,
		firstnumber=2,
    numbersep=3pt,                  
    showspaces=false,                
    showstringspaces=false,
    showtabs=false,                  
    tabsize=2,
		aboveskip=-1pt,
    belowskip=-3pt,
    language=C
}

\lstdefinestyle{CStyle2B2}{
    backgroundcolor=\color{backgroundColourBlue},   
    commentstyle=\color{mGreen},
    keywordstyle=\color{magenta},
    numberstyle=\color{mGray},
    stringstyle=\color{mPurple},
    basicstyle=\tt\fontsize{6.5}{7.5}\selectfont,
    breakatwhitespace=false,         
    breaklines=true,                 
    captionpos=b,
		keepspaces=true,                 
    numbers=left,
    stepnumber=1,
		firstnumber=5,
    numbersep=3pt,                  
    showspaces=false,                
    showstringspaces=false,
    showtabs=false,                  
    tabsize=2,
    belowskip=-3pt,
    language=C
}

\lstdefinestyle{CStyle2B2X}{
    backgroundcolor=\color{backgroundColourBlue},   
    commentstyle=\color{mGreen},
    keywordstyle=\color{magenta},
    numberstyle=\color{mGray},
    stringstyle=\color{mPurple},
    basicstyle=\tt\fontsize{6.5}{7.5}\selectfont,
    breakatwhitespace=false,         
    breaklines=true,                 
    captionpos=b,
		keepspaces=true,                 
    numbers=left,
    stepnumber=1,
		firstnumber=5,
    numbersep=3pt,                  
    showspaces=false,                
    showstringspaces=false,
    showtabs=false,                  
    tabsize=2,
    language=C
}

\lstdefinestyle{CStyle2B3}{
    backgroundcolor=\color{backgroundColourGreen},   
    commentstyle=\color{mGreen},
    keywordstyle=\color{magenta},
    numberstyle=\color{mGray},
    stringstyle=\color{mPurple},
    basicstyle=\tt\fontsize{6.5}{7.5}\selectfont,
    breakatwhitespace=false,         
    breaklines=true,                 
    captionpos=b,
		keepspaces=true,                 
    numbers=left,
    stepnumber=1,
		firstnumber=7,
    numbersep=3pt,                  
    showspaces=false,                
    showstringspaces=false,
    showtabs=false,                  
    tabsize=2,
		aboveskip=-3pt,
    belowskip=-3pt,
    language=C
}

\lstdefinestyle{CStyle2Blast}{
    backgroundcolor=\color{backgroundColour},   
    commentstyle=\color{mGreen},
    keywordstyle=\color{magenta},
    numberstyle=\color{mGray},
    stringstyle=\color{mPurple},
    basicstyle=\tt\fontsize{6.5}{7.5}\selectfont,
    breakatwhitespace=false,         
    breaklines=true,                 
    captionpos=b,
		keepspaces=true,                 
    numbers=left,
    stepnumber=1,
		firstnumber=8,
    numbersep=3pt,                  
    showspaces=false,                
    showstringspaces=false,
    showtabs=false,                  
    tabsize=2,
    language=C
}

\newcommand{\eg}{e.g.,~}
\newcommand{\ie}{i.e.~}
\newcommand{\etc}{etc.~}
\newcommand{\etal}{et al.~}
\newcommand{\vs}{vs~}

\newcommand{\mypara}[1]{\vspace{0.1in}\noindent\textbf{#1}}

\newcommand{\CC}[1]{\textcolor{blue}{[CC: #1]}}
\newcommand{\AZ}[1]{\textcolor{orange}{[AZ: #1]}}

\newcommand{\libprobe}{\textsc{LibProbe}\xspace}
\newcommand{\clang}{\textsc{Clang}\xspace}
\newcommand{\clangversion}{\textsc{11}\xspace}
\newcommand{\clangtidy}{\textsc{clang-tidy}\xspace}
\newcommand{\libtooling}{\textsc{libtooling}\xspace}
\newcommand{\gcov}{\texttt{GCov}\xspace}
\newcommand{\lcov}{\texttt{LCov}\xspace}
\newcommand{\git}{\textsc{Git}\xspace}
\newcommand{\github}{\textsc{GitHub}\xspace}
\newcommand{\grep}{\texttt{Grep}\xspace}
\newcommand{\weggli}{\textsc{Weggli}\xspace}
\newcommand{\libtool}{\textsc{Clang libtooling}\xspace}
\newcommand{\ccscanner}{\textsc{CCScanner}\xspace}
\newcommand{\centris}{\textsc{Centris}\xspace}
\newcommand{\openldap}{\textsc{OpenLDAP}\xspace}
\newcommand{\java}{\textsc{Java}\xspace}
\newcommand{\javascript}{\textsc{JavaScript}\xspace}
\newcommand{\cmake}{\textsc{CMake}\xspace}
\newcommand{\python}{\textsc{Python}\xspace}
\newcommand{\systemd}{\textsc{Systemd}\xspace}

\newcommand{\totalLibs}{21\xspace}
\newcommand{\totalDeps}{51,192\xspace}
\newcommand{\totalClients}{3,198\xspace}
\newcommand{\actualNonUniqueClients}{3,061\xspace}
\newcommand{\actualUniqueClients}{2,070\xspace}
\newcommand{\totalcoverageLibs}{16\xspace}
\newcommand{\ffmpeg}{\textsc{FFmpeg}\xspace}
\newcommand{\fftw}{\textsc{FFTW3}\xspace}
\newcommand{\freetype}{\textsc{FreeType}\xspace}
\newcommand{\glew}{\textsc{GLEW}\xspace}
\newcommand{\glib}{\textsc{GLib}\xspace}
\newcommand{\gsl}{\textsc{GSL}\xspace}
\newcommand{\hdf}{\textsc{HDF5}\xspace}
\newcommand{\hidapi}{\textsc{HIDAPI}\xspace}
\newcommand{\jemalloc}{\textsc{JEMalloc}\xspace}
\newcommand{\lmdb}{\textsc{LMDB}\xspace}
\newcommand{\lz}{\textsc{LZ4}\xspace}
\newcommand{\luajit}{\textsc{LuaJIT}\xspace}
\newcommand{\mbedtls}{\textsc{MbedTLS}\xspace}
\newcommand{\ncurses}{\textsc{NCurses}\xspace}
\newcommand{\ssl}{\textsc{OpenSSL}\xspace}
\newcommand{\sdl}{\textsc{SDL}\xspace}
\newcommand{\sqlite}{\textsc{SQLite}\xspace}
\newcommand{\vorbis}{\textsc{Vorbis}\xspace}
\newcommand{\xxhash}{\textsc{xxHash}\xspace}
\newcommand{\zip}{\textsc{Zip}\xspace}
\newcommand{\zstd}{\textsc{Zstandard}\xspace}

\newcommand{\cava}{\textsc{CAVA}\xspace}
\newcommand{\knot}{\textsc{Knot DNS}\xspace}
\newcommand{\sfml}{\textsc{SFML}\xspace}
\newcommand{\ufo}{\textsc{UFO Alien Invasion}\xspace}
\newcommand{\uacme}{\textsc{uacme}\xspace}
\newcommand{\curl}{\textsc{curl}\xspace}
\newcommand{\openvpn}{\textsc{OpenVPN}\xspace}
\newcommand{\lighttpd}{\textsc{Lighttpd}\xspace}
\newcommand{\libcoap}{\textsc{libCoAP}\xspace}
\newcommand{\krb}{\textsc{Kerberos}\xspace}
\newcommand{\recorder}{\textsc{Recorder}\xspace}
\newcommand{\libetpan}{\textsc{LibEtPan}\xspace}
\newcommand{\fapolicyd}{\textsc{fapolicyd}\xspace}
\newcommand{\osmexpress}{\textsc{OSMExpress}\xspace}

\newcommand{\fftwgit}{https://github.com/FFTW/fftw3\xspace}
\newcommand{\freetypegit}{https://github.com/freetype/freetype\xspace}
\newcommand{\ffmpeggit}{https://github.com/FFmpeg/FFmpeg\xspace}
\newcommand{\glewgit}{https://github.com/nigels-com/glew\xspace}
\newcommand{\glibgit}{https://github.com/GNOME/glib\xspace}
\newcommand{\gslgit}{https://github.com/ampl/gsl\xspace}
\newcommand{\hdfgit}{https://github.com/HDFGroup/hdf5\xspace}
\newcommand{\hidapigit}{https://github.com/libusb/hidapi\xspace}
\newcommand{\jemallocgit}{https://github.com/jemalloc/jemalloc\xspace}
\newcommand{\sqlitegit}{https://github.com/sqlite/sqlite\xspace}
\newcommand{\sslgit}{https://github.com/openssl/openssl\xspace}
\newcommand{\sdlgit}{https://github.com/libsdl-org/SDL\xspace}
\newcommand{\lmdbgit}{https://github.com/LMDB/lmdb\xspace}
\newcommand{\lzgit}{https://github.com/lz4/lz4\xspace}
\newcommand{\luajitgit}{https://github.com/LuaJIT/LuaJIT\xspace}
\newcommand{\mbedtlsgit}{https://github.com/Mbed-TLS/mbedtls\xspace}
\newcommand{\ncursesgit}{https://github.com/mirror/ncurses\xspace}
\newcommand{\vorbisgit}{https://github.com/xiph/vorbis\xspace}
\newcommand{\xxhashgit}{https://github.com/Cyan4973/xxHash\xspace}
\newcommand{\zipgit}{https://github.com/kuba--/zip\xspace}
\newcommand{\zstdgit}{https://github.com/facebook/zstd\xspace}

\newcommand{\ffmpegexportsize}{1037\xspace}
\newcommand{\ffmpegapisize}{880\xspace}
\newcommand{\ffmpegsize}{480,871\xspace}

\newcommand{\fftwexportsize}{617\xspace}
\newcommand{\fftwapisize}{66\xspace}
\newcommand{\fftwsize}{59,846\xspace}
\newcommand{\freetypeexportsize}{220\xspace}
\newcommand{\freetypeapisize}{219}
\newcommand{\freetypesize}{82,996}
\newcommand{\glewexportsize}{9\xspace}
\newcommand{\glewapisize}{9\xspace}
\newcommand{\glewsize}{22,626\xspace}
\newcommand{\glibexportsize}{4,454\xspace}
\newcommand{\glibapisize}{4,417\xspace}
\newcommand{\glibsize}{213,547\xspace}
\newcommand{\gslexportsize}{5,254\xspace}
\newcommand{\gslapisize}{5,222\xspace}
\newcommand{\gslsize}{122,985\xspace}
\newcommand{\hdfexportsize}{3,331\xspace}
\newcommand{\hdfapisize}{983\xspace}
\newcommand{\hdfsize}{344,814\xspace}
\newcommand{\hidapiexportsize}{25\xspace}
\newcommand{\hidapiapisize}{24\xspace}
\newcommand{\hidapisize}{1,351\xspace}
\newcommand{\jemallocexportsize}{24\xspace}
\newcommand{\jemallocapisize}{24\xspace}
\newcommand{\jemallocsize}{16,993\xspace}
\newcommand{\lmdbexportsize}{70\xspace}
\newcommand{\lmdbapisize}{56\xspace}
\newcommand{\lmdbsize}{4,469\xspace}
\newcommand{\lzexportsize}{100\xspace}
\newcommand{\lzapisize}{100\xspace}
\newcommand{\lzsize}{5,934\xspace}
\newcommand{\luajitexportsize}{148\xspace}
\newcommand{\luajitapisize}{148\xspace}
\newcommand{\luajitsize}{6,671\xspace}
\newcommand{\mbedtlsexportsize}{1,232\xspace}
\newcommand{\mbedtlsapisize}{885\xspace}
\newcommand{\mbedtlssize}{36,437\xspace}
\newcommand{\ncursesexportsize}{748\xspace}
\newcommand{\ncursesapisize}{499\xspace}
\newcommand{\ncursessize}{14,716\xspace}
\newcommand{\sslexportsize}{6,240\xspace}
\newcommand{\sslapisize}{5,282\xspace}
\newcommand{\sslsize}{461,116\xspace}
\newcommand{\sdlexportsize}{841\xspace}
\newcommand{\sdlapisize}{838\xspace}
\newcommand{\sdlsize}{56,458\xspace}
\newcommand{\sqliteexportsize}{269\xspace}
\newcommand{\sqliteapisize}{269\xspace}
\newcommand{\sqlitesize}{211,058\xspace}
\newcommand{\vorbisexportsize}{149\xspace}
\newcommand{\vorbisapisize}{78\xspace}
\newcommand{\vorbissize}{6,688\xspace}
\newcommand{\xxhashexportsize}{49\xspace}
\newcommand{\xxhashapisize}{49\xspace}
\newcommand{\xxhashsize}{2,328\xspace}
\newcommand{\zipexportsize}{37\xspace}
\newcommand{\zipapisize}{37\xspace}
\newcommand{\zipsize}{4,909\xspace}
\newcommand{\zstdexportsize}{186\xspace}
\newcommand{\zstdapisize}{186\xspace}
\newcommand{\zstdsize}{24,254\xspace}

\newcommand{\fftwlib}{\textsc{libfftw3.so}\xspace}
\newcommand{\freetypelib}{\textsc{libfreetype.so}\xspace}
\newcommand{\ffmpeglibs}{\textsc{libavdevice.so, libavfilter.so, libavutil.so, libavcodec.so, libswscale.so, libavformat.so, libswresample.so}\xspace}
\newcommand{\glewlib}{\textsc{libGLEW.so}\xspace}
\newcommand{\gliblibs}{\textsc{libglib-2.0.so, libgthread-2.0.so, libgobject-2.0.so, libgmodule-2.0.so, libgio-2.0.so}\xspace}
\newcommand{\hdflib}{\textsc{libhdf5.so, libhdf5\_hl.so}\xspace}
\newcommand{\hidapilibs}{\textsc{libhidapi-libusb.so, libhidapi-hidraw.so}\xspace}
\newcommand{\jemalloclib}{\textsc{libjemalloc.so}\xspace}
\newcommand{\sqlitelib}{\textsc{libsqlite3.so}\xspace}
\newcommand{\ssllib}{\textsc{libssl.so, libcrypto.so}\xspace}
\newcommand{\sdllib}{\textsc{libSDL2.so}\xspace}
\newcommand{\lmdblib}{\textsc{liblmdb.so}\xspace}
\newcommand{\lzlib}{\textsc{liblz4.so}\xspace}
\newcommand{\luajitlib}{\textsc{liblua.so}\xspace}
\newcommand{\mbedtlslibs}{\textsc{libmbedcrypto.so, libmbedtls.so, libmbedx509.so}\xspace}
\newcommand{\ncurseslibs}{\textsc{libncurses.so, libform.so, libpanel.so}\xspace}
\newcommand{\vorbislibs}{\textsc{libvorbis.so, libvorbisenc.so, libvorbisfile.so}\xspace}
\newcommand{\xxhashlib}{\textsc{libxxhash.so}\xspace}
\newcommand{\ziplib}{\textsc{libzip.so}\xspace}
\newcommand{\zstdlib}{\textsc{libzstd.so}\xspace}
\newcommand{\apisize}{API implementation\xspace}

\newcommand{\fftwclients}{151}
\newcommand{\fftwuclients}{65}
\newcommand{\freetypeclients}{355}
\newcommand{\freetypeuclients}{263}
\newcommand{\ffmpegclients}{236}
\newcommand{\ffmpeguclients}{120}
\newcommand{\glewclients}{490}
\newcommand{\glewuclients}{387}
\newcommand{\glibclients}{365}
\newcommand{\glibuclients}{302}
\newcommand{\gslclients}{162}
\newcommand{\gsluclients}{119}
\newcommand{\hdfclients}{215}
\newcommand{\hdfuclients}{159}
\newcommand{\hidapiclients}{100}
\newcommand{\hidapiuclients}{83}
\newcommand{\jemallocclients}{159}
\newcommand{\jemallocuclients}{144}
\newcommand{\sqliteclients}{428}
\newcommand{\sqliteuclients}{143}
\newcommand{\sslclients}{444}
\newcommand{\ssluclients}{131}
\newcommand{\sdlclients}{196}
\newcommand{\sdluclients}{70}
\newcommand{\lmdbclients}{131}
\newcommand{\lmdbuclients}{115}
\newcommand{\luajitclients}{284}
\newcommand{\luajituclients}{256}
\newcommand{\lzclients}{498}
\newcommand{\lzuclients}{31}
\newcommand{\mbedtlsclients}{179}
\newcommand{\mbedtlsuclients}{134}
\newcommand{\ncursesclients}{156}
\newcommand{\ncursesuclients}{136}
\newcommand{\vorbisclients}{165}
\newcommand{\vorbisuclients}{130}
\newcommand{\xxhashclients}{343}
\newcommand{\xxhashuclients}{120}
\newcommand{\zipclients}{109}
\newcommand{\zipuclients}{81}
\newcommand{\zstdclients}{189}
\newcommand{\zstduclients}{72}

\author{Ahmed Zaki}
\email{ahmed.zaki@imperial.ac.uk}
\orcid{0009-0008-7141-8865}
\affiliation{
  \institution{Imperial College London}
  \city{London}
  \country{UK}
}

\author{Cristian Cadar}
\email{c.cadar@imperial.ac.uk}
\orcid{0000-0002-3599-7264}
\affiliation{
  \institution{Imperial College London}
  \city{London}
  \country{UK}
}

\newcommand{\libsWithUnusedApis}{16\xspace}

\newcommand{\unusedApiFFmpeg}{221\xspace}
\newcommand{\ffmpegApiUsagePercentage}{75\%\xspace}
\newcommand{\ffmpegApiUnusedPercentage}{25\%\xspace}

\newcommand{\unusedApiFftw}{33\xspace}
\newcommand{\fftwApiUsagePercentage}{50\%\xspace}
\newcommand{\fftwApiUnusedPercentage}{50\%\xspace}

\newcommand{\unusedApiFreetype}{17\xspace}
\newcommand{\freetypeApiUsagePercentage}{92\%\xspace}
\newcommand{\freetypeApiUnusedPercentage}{8\%\xspace}

\newcommand{\unusedApiGlew}{0\xspace}
\newcommand{\glewApiUsagePercentage}{100\%\xspace}
\newcommand{\glewApiUnusedPercentage}{0\%\xspace}

\newcommand{\unusedApiGlib}{1,521\xspace}
\newcommand{\glibApiUsagePercentage}{66\%\xspace}
\newcommand{\glibApiUnusedPercentage}{34\%\xspace}

\newcommand{\unusedApiGsl}{2,459\xspace}
\newcommand{\gslApiUsagePercentage}{47\%\xspace}
\newcommand{\gslApiUnusedPercentage}{53\%\xspace}

\newcommand{\unusedApiHdf}{603\xspace}
\newcommand{\hdfApiUsagePercentage}{39\%\xspace}
\newcommand{\hdfApiUnusedPercentage}{61\%\xspace}

\newcommand{\unusedApiHidapi}{0\xspace}
\newcommand{\hidapiApiUsagePercentage}{100\%\xspace}
\newcommand{\hidapiApiUnusedPercentage}{0\%\xspace}

\newcommand{\unusedApiJemalloc}{3\xspace}
\newcommand{\jemallocApiUsagePercentage}{87\%\xspace}
\newcommand{\jemallocApiUnusedPercentage}{13\%\xspace}

\newcommand{\unusedApiLmdb}{10\xspace}
\newcommand{\lmdbApiUsagePercentage}{82\%\xspace}
\newcommand{\lmdbApiUnusedPercentage}{18\%\xspace}

\newcommand{\unusedApiLuajit}{2\xspace}
\newcommand{\luajitApiUsagePercentage}{99\%\xspace}
\newcommand{\luajitApiUnusedPercentage}{1\%\xspace}

\newcommand{\unusedApiLz}{17\xspace}
\newcommand{\lzApiUsagePercentage}{83\%\xspace}
\newcommand{\lzApiUnusedPercentage}{17\%\xspace}

\newcommand{\unusedApiMbedtls}{96\xspace}
\newcommand{\mbedtlsApiUsagePercentage}{89\%\xspace}
\newcommand{\mbedtlsApiUnusedPercentage}{11\%\xspace}

\newcommand{\unusedApiNcurses}{12\xspace}
\newcommand{\ncursesApiUsagePercentage}{98\%\xspace}
\newcommand{\ncursesApiUnusedPercentage}{2\%\xspace}

\newcommand{\unusedApiSdl}{432\xspace}
\newcommand{\sdlApiUsagePercentage}{48\%\xspace}
\newcommand{\sdlApiUnusedPercentage}{52\%\xspace}

\newcommand{\unusedApiSsl}{274\xspace}
\newcommand{\sslApiUsagePercentage}{95\%\xspace}
\newcommand{\sslApiUnusedPercentage}{5\%\xspace}

\newcommand{\unusedApiSqlite}{0\xspace}
\newcommand{\sqliteApiUsagePercentage}{100\%\xspace}
\newcommand{\sqliteApiUnusedPercentage}{0\%\xspace}

\newcommand{\unusedApiVorbis}{0\xspace}
\newcommand{\vorbisApiUsagePercentage}{100\%\xspace}
\newcommand{\vorbisApiUnusedPercentage}{0\%\xspace}

\newcommand{\unusedApiXxhash}{0\xspace}
\newcommand{\xxhashApiUsagePercentage}{100\%\xspace}
\newcommand{\xxhashApiUnusedPercentage}{0\%\xspace}

\newcommand{\unusedApiZip}{14\xspace}
\newcommand{\zipApiUsagePercentage}{62\%\xspace}
\newcommand{\zipApiUnusedPercentage}{38\%\xspace}

\newcommand{\unusedApiZstd}{1\xspace}
\newcommand{\zstdApiUsagePercentage}{99\%\xspace}
\newcommand{\zstdApiUnusedPercentage}{1\%\xspace}

\newcommand{\ffmpegmostinvoked}{\textit{av\_log}\xspace}
\newcommand{\ffmpeginvocations}{15,478\xspace}
\newcommand{\fftwmostinvoked}{\textit{fftw\_destroy\_plan}\xspace}
\newcommand{\fftwinvocations}{196\xspace}
\newcommand{\freetypemostinvoked}{\textit{FT\_Get\_Char\_Index}\xspace}
\newcommand{\freetypeinvocations}{667\xspace}
\newcommand{\hdfmostinvoked}{\textit{H5Sclose}\xspace}
\newcommand{\hdfinvocations}{1553\xspace}
\newcommand{\hidapimostinvoked}{\textit{hid\_write}\xspace}
\newcommand{\hidapiinvocations}{912\xspace}
\newcommand{\jemallocmostinvoked}{\textit{free}\xspace}
\newcommand{\jemallocinvocations}{213,094\xspace}
\newcommand{\lmdbmostinvoked}{\textit{mdb\_strerror}\xspace}
\newcommand{\lmdbinvocations}{513\xspace}
\newcommand{\lzmostinvoked}{\textit{LZ4F\_isError}\xspace}
\newcommand{\lzinvocations}{58\xspace}
\newcommand{\luajitmostinvoked}{\textit{lua\_gettop}\xspace}
\newcommand{\luajitinvocations}{45,819\xspace}
\newcommand{\mbedtlsmostinvoked}{\textit{mbedtls\_mpi\_free}\xspace}
\newcommand{\mbedtlsinvocations}{1782\xspace}
\newcommand{\ncursesmostinvoked}{\textit{move}\xspace}
\newcommand{\ncursesinvocations}{42,571\xspace}
\newcommand{\vorbismostinvoked}{\textit{ov\_clear}\xspace}
\newcommand{\vorbisinvocations}{274\xspace}
\newcommand{\xxhashmostinvoked}{\textit{XXH64}\xspace}
\newcommand{\xxhashinvocations}{570\xspace}
\newcommand{\zipinvocations}{247\xspace}
\newcommand{\zipmostinvoked}{\textit{zip\_close}\xspace}
\newcommand{\zstdinvocations}{1,615\xspace}
\newcommand{\zstdmostinvoked}{\textit{ZSTD\_isError}\xspace}
\newcommand{\sqliteinvocations}{9,7505\xspace}
\newcommand{\sqlitemostinvoked}{\textit{sqlite3\_free}\xspace}
\newcommand{\sslinvocations}{2,846\xspace}
\newcommand{\sslmostinvoked}{\textit{BIO\_ctrl}\xspace}
\newcommand{\sdlinvocations}{875\xspace}
\newcommand{\sdlmostinvoked}{\textit{SDL\_GetError}\xspace}
\newcommand{\glewinvocations}{903\xspace}
\newcommand{\glewmostinvoked}{\textit{glewInit}\xspace}
\newcommand{\glibinvocations}{9,3564\xspace}
\newcommand{\glibmostinvoked}{\textit{g\_free}\xspace}
\newcommand{\gslinvocations}{890\xspace}
\newcommand{\gslmostinvoked}{\textit{gsl\_test}\xspace}
\newcommand{\mbedtlstopclientname}{\textit{cosmopolitan}\xspace}
\newcommand{\mbedtlstopclientcount}{11,039\xspace}
\newcommand{\fftwtopclientname}{\textit{cava}\xspace}
\newcommand{\fftwtopclientcount}{48\xspace}
\newcommand{\lmdbtopclientname}{\textit{ptarmigan}\xspace}
\newcommand{\lmdbtopclientcount}{431\xspace}
\newcommand{\luajittopclientname}{\textit{adxe}\xspace}
\newcommand{\luajittopclientcount}{17,147\xspace}
\newcommand{\jemalloctopclientname}{\textit{airyx}\xspace}
\newcommand{\jemalloctopclientcount}{65,820\xspace}
\newcommand{\hdftopclientname}{\textit{ohpc}\xspace}
\newcommand{\hdftopclientcount}{4,354\xspace}
\newcommand{\freetypetopclientname}{\textit{SDL\_gui}\xspace}
\newcommand{\freetypetopclientcount}{1,403}
\newcommand{\ziptopclientname}{\textit{Torque3D}\xspace}
\newcommand{\ziptopclientcount}{142\xspace}
\newcommand{\hidapitopclientname}{\textit{OpenRGB}\xspace}
\newcommand{\hidapitopclientcount}{1667\xspace}
\newcommand{\ncursestopclientname}{\textit{ceph}\xspace}
\newcommand{\ncursestopclientcount}{12655\xspace}
\newcommand{\vorbistopclientname}{\textit{darkice}\xspace}
\newcommand{\vorbistopclientcount}{741\xspace}
\newcommand{\xxhashtopclientname}{\textit{qs}\xspace}
\newcommand{\xxhashtopclientcount}{142\xspace}
\newcommand{\glewtopclientname}{\textit{OGLdev}\xspace}
\newcommand{\glewtopclientcount}{84\xspace}
\newcommand{\ssltopclientname}{\textit{ApertusVR}\xspace}
\newcommand{\ssltopclientcount}{45,171\xspace}
\newcommand{\sqlitetopclientname}{\textit{Gideros}\xspace}
\newcommand{\sqlitetopclientcount}{19328\xspace}

\newcommand{\totalapislessthantwenty}{13,787\xspace}
\newcommand{\fullcovlessthantwentyapis}{6,738\xspace}
\newcommand{\fullcovlessthantwentyapispercent}{48.8\%\xspace}
\newcommand{\totalineslessthantwenty}{89,203}
\newcommand{\totalineslessthantwentycovered}{61011}
\newcommand{\totalineslessthantwentypercent}{68.3\%\xspace}
\newcommand{\totalapismorethantwenty}{2,120\xspace}
\newcommand{\fullcovmorethantwentyapis}{286\xspace}
\newcommand{\fullcovmorethantwentyapispercent}{13.4\%\xspace}
\newcommand{\totalinesmorethantwenty}{89,993}
\newcommand{\totalinesmorethantwentycovered}{62161}
\newcommand{\totalinesmorethantwentypercent}{69.3\%\xspace}

\newcommand{\lmdbapiwithclientsnocov}{{\textit{mdv\_env\_info}}\xspace}
\newcommand{\lmdbapiwithclientsnocovnum}{61\xspace}
\newcommand{\lmdbapiwithclientsnocovpercent}{53\%\xspace}
\newcommand{\vorbisapiwithclientsnocov}{{\textit{ov\_info}}\xspace}
\newcommand{\vorbisapiwithclientsnocovnum}{106\xspace}
\newcommand{\vorbisapiwithclientsnocovpercent}{81\%\xspace}

\newcommand{\fftwcovA}{36.7\%\xspace}
\newcommand{\fftwcovB}{93.2\%\xspace}
\newcommand{\fftwcovC}{93.1\%\xspace}
\newcommand{\fftwcovD}{40.7\%\xspace}
\newcommand{\fftwcovE}{51.7\%\xspace}
\newcommand{\fftwcovF}{78.6\%\xspace}
\newcommand{\fftwcovG}{51.5\%\xspace}
\newcommand{\fftwcovH}{53.2\%\xspace}
\newcommand{\fftwcovI}{89.9\%\xspace}
\newcommand{\fftwcov}{51.7\%\xspace}
\newcommand{\fftwcovmin}{36.7\%\xspace}
\newcommand{\fftwcovmax}{93.2\%\xspace}

\newcommand{\ffmpegcov}{60.5\%\xspace}
\newcommand{\ffmpegapislessthantwenty}{703\xspace}
\newcommand{\ffmpegapislessthantwentycov}{78.2\%\xspace}
\newcommand{\ffmpegapismorethantwenty}{143\xspace}
\newcommand{\ffmpegapismorethantwentycov}{74.3\%\xspace}
\newcommand{\ffmpegusedanduncoveredapis}{128\xspace}
\newcommand{\ffmpegusedanduncoveredapispercent}{14\%\xspace}

\newcommand{\fftwusedanduncoveredapis}{14\xspace}
\newcommand{\fftwusedanduncoveredapispercent}{21\%\xspace}

\newcommand{\glibcov}{73.3\%\xspace}
\newcommand{\glibapislessthantwenty}{3892\xspace}
\newcommand{\glibapislessthantwentycov}{79.6\%\xspace}
\newcommand{\glibapismorethantwenty}{328\xspace}
\newcommand{\glibapismorethantwentycov}{78.6\%\xspace}
\newcommand{\glibusedanduncoveredapis}{335\xspace}
\newcommand{\glibusedanduncoveredapispercent}{7\%\xspace}

\newcommand{\gslcov}{86.3\%\xspace}
\newcommand{\gslapislessthantwenty}{1729\xspace}
\newcommand{\gslapislessthantwentycov}{69.8\%\xspace}
\newcommand{\gslapismorethantwenty}{407\xspace}
\newcommand{\gslapismorethantwentycov}{75.5\%\xspace}
\newcommand{\gslusedanduncoveredapis}{262\xspace}
\newcommand{\gslusedanduncoveredapispercent}{5\%\xspace}

\newcommand{\jemalloccov}{84.5\%\xspace}
\newcommand{\jemallocapislessthantwenty}{16\xspace}
\newcommand{\jemallocapislessthantwentycov}{63.5\%\xspace}
\newcommand{\jemallocapismorethantwenty}{8\xspace}
\newcommand{\jemallocapismorethantwentycov}{78.3\%\xspace}
\newcommand{\jemallocusedanduncoveredapis}{4\xspace}
\newcommand{\jemallocusedanduncoveredapispercent}{17\%\xspace}

\newcommand{\lzcov}{70.4\%\xspace}
\newcommand{\lzapislessthantwenty}{93\xspace}
\newcommand{\lzapislessthantwentycov}{63.1\%\xspace}
\newcommand{\lzapismorethantwenty}{7\xspace}
\newcommand{\lzapismorethantwentycov}{44.2\%\xspace}
\newcommand{\lzusedanduncoveredapis}{45\xspace}
\newcommand{\lzusedanduncoveredapispercent}{45\%\xspace}

\newcommand{\xxhashcov}{76.0\%\xspace}
\newcommand{\xxhashapislessthantwenty}{47\xspace}
\newcommand{\xxhashapislessthantwentycov}{89.2\%\xspace}
\newcommand{\xxhashapismorethantwenty}{2\xspace}
\newcommand{\xxhashapismorethantwentycov}{100.0\%\xspace}
\newcommand{\xxhashusedanduncoveredapis}{8\xspace}
\newcommand{\xxhashusedanduncoveredapispercent}{16\%\xspace}
\newcommand{\mbedtlscov}{78.9\%\xspace}

\newcommand{\lmdbcov}{38.2\%\xspace}
\newcommand{\lmdbfncov}{57\%\xspace}
\newcommand{\lmdbcovlinenumbers}{(1708 of 4469 lines)\xspace}
\newcommand{\lmdbcovfnnumbers}{(93 of 162 function)\xspace}

\newcommand{\luajitcov}{78.8\%\xspace}
\newcommand{\hdfcov}{69.3\%\xspace}

\newcommand{\zipcov}{42.5\%\xspace}

\newcommand{\vorbiscov}{58.9\%\xspace} 

\newcommand{\sqliteapislessthantwenty}{243\xspace}
\newcommand{\sqliteapislessthantwentycov}{91.7\%\xspace}
\newcommand{\sqliteapismorethantwenty}{26\xspace}
\newcommand{\sqliteapismorethantwentycov}{87.3\%\xspace}
\newcommand{\sqlitecov}{61.0\%\xspace}
\newcommand{\sqliteusedanduncoveredapis}{19\xspace}
\newcommand{\sqliteusedanduncoveredapispercent}{7\%\xspace}

\newcommand{\sslapislessthantwenty}{4520\xspace}
\newcommand{\sslapislessthantwentycov}{59.0\%\xspace}
\newcommand{\sslapismorethantwenty}{622\xspace}
\newcommand{\sslapismorethantwentycov}{64.1\%\xspace}
\newcommand{\sslcov}{63.7\%\xspace}
\newcommand{\sslusedanduncoveredapis}{1560\xspace}
\newcommand{\sslusedanduncoveredapispercent}{29\%\xspace}

\newcommand{\sdlapislessthantwenty}{746\xspace}
\newcommand{\sdlapislessthantwentycov}{42.3\%\xspace}
\newcommand{\sdlapismorethantwenty}{84\xspace}
\newcommand{\sdlapismorethantwentycov}{45.8\%\xspace}
\newcommand{\sdlcov}{18.3\%\xspace}
\newcommand{\sdlusedanduncoveredapis}{182\xspace}
\newcommand{\sdlusedanduncoveredapispercent}{22\%\xspace}

\newcommand{\zstdapislessthantwenty}{166\xspace}
\newcommand{\zstdapislessthantwentycov}{88.2\%\xspace}
\newcommand{\zstdapismorethantwenty}{20\xspace}
\newcommand{\zstdapismorethantwentycov}{84.5\%\xspace}
\newcommand{\zstdcov}{71.1\%\xspace}
\newcommand{\zstdusedanduncoveredapis}{16\xspace}
\newcommand{\zstdusedanduncoveredapispercent}{8\%\xspace}

\newcommand{\fftwuncovered}{14\xspace}
\newcommand{\fftwuseduncovpercentage}{40\%\xspace}
\newcommand{\mbedtlsuncovered}{77\xspace}
\newcommand{\mbedtlsuseduncovpercentage}{10\%\xspace}
\newcommand{\lmdbuncovered}{29\xspace}
\newcommand{\lmdbuseduncovpercentage}{63\%\xspace}
\newcommand{\hdfuncovered}{16\xspace}
\newcommand{\hdfuseduncovpercentage}{4\%\xspace}

\newcommand{\zipUncovered}{a single\xspace}
\newcommand{\zipUncoveredApi}{zip\_entry\_fread\xspace}

\newcommand{\zipuseduncovpercentage}{4\%\xspace}
\newcommand{\vorbisuncovered}{51\xspace}
\newcommand{\vorbisuseduncovpercentage}{64\%\xspace}

\newcommand{\xxhashuncovered}{8\xspace}

\newcommand{\mbedtlsapislessthantwenty}{663\xspace}
\newcommand{\mbedtlsapislessthantwentycov}{83.2\%\xspace}
\newcommand{\mbedtlsapismorethantwenty}{222\xspace}
\newcommand{\mbedtlsapismorethantwentycov}{80.6\%}
\newcommand{\mbedtlsusedanduncoveredapis}{87\xspace}
\newcommand{\mbedtlsusedanduncoveredapispercent}{10\%\xspace}

\newcommand{\fftwapislessthantwenty}{66\xspace}
\newcommand{\fftwapislessthantwentycov}{43.3\%\xspace}
\newcommand{\fftwapismorethantwenty}{0\xspace}
\newcommand{\fftwapismorethantwentycov}{-\xspace}

\newcommand{\lmdbapislessthantwenty}{49\xspace}
\newcommand{\lmdbapislessthantwentycov}{28.1\%\xspace}
\newcommand{\lmdbapismorethantwenty}{7\xspace}
\newcommand{\lmdbapismorethantwentycov}{25.6\%\xspace}
\newcommand{\lmdbusedanduncoveredapis}{25\xspace}
\newcommand{\lmdbusedanduncoveredapispercent}{45\%\xspace}

\newcommand{\zipapislessthantwenty}{29\xspace}
\newcommand{\zipapislessthantwentycov}{81.6\%\xspace}
\newcommand{\zipapismorethantwenty}{8\xspace}
\newcommand{\zipapismorethantwentycov}{63.1\%\xspace}
\newcommand{\zipusedanduncoveredapis}{1\xspace}
\newcommand{\zipusedanduncoveredapispercent}{3\%\xspace}
\newcommand{\vorbisapislessthantwenty}{59\xspace}
\newcommand{\vorbisapislessthantwentycov}{33.7\%\xspace}
\newcommand{\vorbisapismorethantwenty}{19\xspace}
\newcommand{\vorbisapismorethantwentycov}{36.8\%\xspace}

\newcommand{\hdfapislessthantwenty}{766\xspace}
\newcommand{\hdfapislessthantwentycov}{61.7\%\xspace}
\newcommand{\hdfapismorethantwenty}{217\xspace}
\newcommand{\hdfapismorethantwentycov}{61.8\%\xspace}
\newcommand{\hdfusedanduncoveredapis}{36\xspace}
\newcommand{\hdfusedanduncoveredapispercent}{4\%\xspace}

\newcommand{\lmdblinecovafterknot}{53.5\%\xspace}
\newcommand{\lmdbfunctioncovafterknot}{64.8\%\xspace}


\newcommand{\lmdbimprovedcovapis}{4\xspace}
\newcommand{\lmdbnewlycoveredapis}{4\xspace}
\newcommand{\lmdbnewlycoveredlinesapis}{123\xspace}
\newcommand{\lmdbimprovedcov}{14.7\%\xspace}

\newcommand{\lmdbnewlycoveredlinesapispercent}{17.35\%\xspace}
\newcommand{\vorbisimprovedcovpis}{1\xspace}
\newcommand{\vorbisnewlycoveredapis}{14\xspace}
\newcommand{\vorbisnewlycoveredlinesapis}{247\xspace}
\newcommand{\vorbisnewlycoveredlinesapispercent}{15.38\%\xspace}
\newcommand{\vorbistotallines}{6688\xspace}
\newcommand{\vorbislinecovimproved}{66.60\%\xspace}
\newcommand{\vorbisbaselinelinecov}{3942\xspace}
\newcommand{\vorbisimprovedlinecov}{4454\xspace}
\newcommand{\vorbisusedanduncoveredapis}{51\xspace}
\newcommand{\vorbisusedanduncoveredapispercent}{65\%\xspace}

\newcommand{\vorbisdeltaaddedcov}{512\xspace}
\newcommand{\vorbisimprovedcov}{7.7\%\xspace} 
\newcommand{\fftwnewlinecov}{42\%\xspace}
\newcommand{\fftwimprovedcovapis}{0\xspace}
\newcommand{\fftwnewlycoveredapis}{2\xspace}
\newcommand{\fftwnewlycoveredlinesapis}{4\xspace}
\newcommand{\fftwnewlycoveredlinesapispercent}{0.16\%\xspace}
\newcommand{\fftwimprovedcov}{2\%\xspace}
\newcommand{\fftwmaximumlinecovimprove}{8,188\xspace}
\newcommand{\fftwapione}{\textit{fftw\_alloc\_real}\xspace}
\newcommand{\fftwapitwo}{\textit{fftw\_alloc\_complex}\xspace}

\newcommand{\sdlnewlycoveredapis}{6\xspace}
\newcommand{\sdlimprovedcovapis}{0\xspace}
\newcommand{\sdlimprovedcovlines}{89\xspace}
\newcommand{\sdlimprovedcov}{0.3\%\xspace}
\newcommand{\sdlnewlycoveredlinesapispercent}{0.57\%\xspace}
\newcommand{\sdlnewlycoveredlinesapis}{46\xspace}


\newcommand{\grepApiRecallUacme}{1.00}
\newcommand{\grepApiPrecisionUacme}{0.94}
\newcommand{\grepUsageRecallUacme}{0.99}
\newcommand{\grepUsagePrecisionUacme}{0.95}

\newcommand{\weggliApiRecallUacme}{0.97}
\newcommand{\weggliApiPrecisionUacme}{0.92}
\newcommand{\weggliUsageRecallUacme}{0.97}
\newcommand{\weggliUsagePrecisionUacme}{0.90}

\newcommand{\grepApiRecallCurl}{1.00}
\newcommand{\grepApiPrecisionCurl}{0.93}
\newcommand{\grepUsageRecallCurl}{1.00}
\newcommand{\grepUsagePrecisionCurl}{0.81}

\newcommand{\weggliApiRecallCurl}{0.86}
\newcommand{\weggliApiPrecisionCurl}{0.92}
\newcommand{\weggliUsageRecallCurl}{0.84}
\newcommand{\weggliUsagePrecisionCurl}{0.80}

\newcommand{\grepApiRecallLibssh}{1.00}
\newcommand{\grepApiPrecisionLibssh}{1.00}
\newcommand{\grepUsageRecallLibssh}{0.30}
\newcommand{\grepUsagePrecisionLibssh}{0.90}

\newcommand{\weggliApiRecallLibssh}{0.92}
\newcommand{\weggliApiPrecisionLibssh}{0.98}
\newcommand{\weggliUsageRecallLibssh}{0.29}
\newcommand{\weggliUsagePrecisionLibssh}{0.81}

\newcommand{\grepApiRecallOpenvpn}{1.00}
\newcommand{\grepApiPrecisionOpenvpn}{0.92}
\newcommand{\grepUsageRecallOpenvpn}{1.00}
\newcommand{\grepUsagePrecisionOpenvpn}{0.87}

\newcommand{\weggliApiRecallOpenvpn}{1.00}
\newcommand{\weggliApiPrecisionOpenvpn}{0.91}
\newcommand{\weggliUsageRecallOpenvpn}{1.00}
\newcommand{\weggliUsagePrecisionOpenvpn}{0.85}

\newcommand{\grepApiRecallLighttpd}{0.98}
\newcommand{\grepApiPrecisionLighttpd}{0.63}
\newcommand{\grepUsageRecallLighttpd}{0.92}
\newcommand{\grepUsagePrecisionLighttpd}{0.57}

\newcommand{\weggliApiRecallLighttpd}{1.00}
\newcommand{\weggliApiPrecisionLighttpd}{0.64}
\newcommand{\weggliUsageRecallLighttpd}{1.00}
\newcommand{\weggliUsagePrecisionLighttpd}{0.58}

\newcommand{\grepApiRecallLibcoap}{1.00}
\newcommand{\grepApiPrecisionLibcoap}{0.98}
\newcommand{\grepUsageRecallLibcoap}{1.00}
\newcommand{\grepUsagePrecisionLibcoap}{0.96}

\newcommand{\weggliApiRecallLibcoap}{0.95}
\newcommand{\weggliApiPrecisionLibcoap}{0.92}
\newcommand{\weggliUsageRecallLibcoap}{0.90}
\newcommand{\weggliUsagePrecisionLibcoap}{0.89}

\newcommand{\grepTotalDistinctAPIUacme}{219\xspace}
\newcommand{\weggliTotalDistinctAPIUacme}{228\xspace}
\newcommand{\grepTotalDistinctAPICurl}{119\xspace}
\newcommand{\weggliTotalDistinctAPICurl}{106\xspace}
\newcommand{\grepTotalDistinctAPILibssh}{149\xspace}
\newcommand{\weggliTotalDistinctAPILibssh}{147\xspace}

\newcommand{\totalgrepmbedtlsuacme}{219}
\newcommand{\totalwegglimbedtlsuacme}{228}
\newcommand{\totallibtoolmbedtlsuacme}{227}
\newcommand{\missedgrepmbedtlsuacme}{1}
\newcommand{\missedwegglimbedtlsuacme}{5}
\newcommand{\ingrepandlibtoolmbedtlsuacme}{211}
\newcommand{\inweggliandlibtoolmbedtlsuacme}{207}
\newcommand{\totalgrepmbedtlslibzip}{26}
\newcommand{\totalwegglimbedtlslibzip}{21}
\newcommand{\totallibtoolmbedtlslibzip}{23}
\newcommand{\missedgrepmbedtlslibzip}{0}
\newcommand{\missedwegglimbedtlslibzip}{3}
\newcommand{\ingrepandlibtoolmbedtlslibzip}{23}
\newcommand{\inweggliandlibtoolmbedtlslibzip}{20}
\newcommand{\totalgrepmbedtlslibssh}{148}
\newcommand{\totalwegglimbedtlslibssh}{147}
\newcommand{\totallibtoolmbedtlslibssh}{248}
\newcommand{\missedgrepmbedtlslibssh}{229}
\newcommand{\missedwegglimbedtlslibssh}{216}
\newcommand{\ingrepandlibtoolmbedtlslibssh}{32}
\newcommand{\inweggliandlibtoolmbedtlslibssh}{19}
\newcommand{\totalgrepmbedtlscurl}{109}
\newcommand{\totalwegglimbedtlscurl}{96}
\newcommand{\totallibtoolmbedtlscurl}{92}
\newcommand{\missedgrepmbedtlscurl}{10}
\newcommand{\missedwegglimbedtlscurl}{0}
\newcommand{\ingrepandlibtoolmbedtlscurl}{92}
\newcommand{\inweggliandlibtoolmbedtlscurl}{82}

\newcommand{\tbltotalgrepmbedtlsuacme}{103}
\newcommand{\tbltotalwegglimbedtlsuacme}{103}
\newcommand{\tbltotallibtoolmbedtlsuacme}{106}
\newcommand{\tblingrepandlibtoolmbedtlsuacme}{95}
\newcommand{\tblinweggliandlibtoolmbedtlsuacme}{97}
\newcommand{\tbltotalgrepmbedtlslibzip}{18}
\newcommand{\tbltotalwegglimbedtlslibzip}{17}
\newcommand{\tbltotallibtoolmbedtlslibzip}{18}
\newcommand{\tblingrepandlibtoolmbedtlslibzip}{18}
\newcommand{\tblinweggliandlibtoolmbedtlslibzip}{16}
\newcommand{\tbltotalgrepmbedtlslibssh}{66}
\newcommand{\tbltotalwegglimbedtlslibssh}{63}
\newcommand{\tbltotallibtoolmbedtlslibssh}{9}
\newcommand{\tblingrepandlibtoolmbedtlslibssh}{9}
\newcommand{\tblinweggliandlibtoolmbedtlslibssh}{4}
\newcommand{\tbltotalgrepmbedtlscurl}{70}
\newcommand{\tbltotalwegglimbedtlscurl}{58}
\newcommand{\tbltotallibtoolmbedtlscurl}{66}
\newcommand{\tblingrepandlibtoolmbedtlscurl}{66}
\newcommand{\tblinweggliandlibtoolmbedtlscurl}{58}

\newcommand{\grepApiRecallKrb}{1.00}
\newcommand{\grepApiPrecisionKrb}{1.00}
\newcommand{\grepUsageRecallKrb}{1.00}
\newcommand{\grepUsagePrecisionKrb}{1.00}

\newcommand{\weggliApiRecallKrb}{1.00}
\newcommand{\weggliApiPrecisionKrb}{1.00}
\newcommand{\weggliUsageRecallKrb}{1.00}
\newcommand{\weggliUsagePrecisionKrb}{1.00}

\newcommand{\grepApiRecallRecorder}{1.00}
\newcommand{\grepApiPrecisionRecorder}{1.00}
\newcommand{\grepUsageRecallRecorder}{1.00}
\newcommand{\grepUsagePrecisionRecorder}{1.00}

\newcommand{\weggliApiRecallRecorder}{1.00}
\newcommand{\weggliApiPrecisionRecorder}{1.00}
\newcommand{\weggliUsageRecallRecorder}{1.00}
\newcommand{\weggliUsagePrecisionRecorder}{1.00}

\newcommand{\grepApiRecallKnot}{1.00}
\newcommand{\grepApiPrecisionKnot}{1.00}
\newcommand{\grepUsageRecallKnot}{1.00}
\newcommand{\grepUsagePrecisionKnot}{0.98}

\newcommand{\weggliApiRecallKnot}{1.00}
\newcommand{\weggliApiPrecisionKnot}{1.00}

\newcommand{\grepApiRecallLibetpan}{0.93}
\newcommand{\grepApiPrecisionLibetpan}{1.00}
\newcommand{\grepUsageRecallLibetpan}{0.97}
\newcommand{\grepUsagePrecisionLibetpan}{1.00}

\newcommand{\weggliApiRecallLibetpan}{1.00}
\newcommand{\weggliApiPrecisionLibetpan}{1.00}
\newcommand{\weggliUsageRecallLibetpan}{1.00}
\newcommand{\weggliUsagePrecisionLibetpan}{1.00}

\newcommand{\grepApiRecallFapolicyd}{0.95}
\newcommand{\grepApiPrecisionFapolicyd}{0.95}
\newcommand{\grepUsageRecallFapolicyd}{0.96}
\newcommand{\grepUsagePrecisionFapolicyd}{0.95}

\newcommand{\weggliApiRecallFapolicyd}{0.95}
\newcommand{\weggliApiPrecisionFapolicyd}{1.00}
\newcommand{\weggliUsageRecallFapolicyd}{0.97}
\newcommand{\weggliUsagePrecisionFapolicyd}{1.00}

\newcommand{\grepApiRecallOsmexpress}{1.00}
\newcommand{\grepApiPrecisionOsmexpress}{1.00}
\newcommand{\grepUsageRecallOsmexpress}{0.89}
\newcommand{\grepUsagePrecisionOsmexpress}{0.81}

\newcommand{\weggliApiRecallOsmexpress}{0.92}
\newcommand{\weggliApiPrecisionOsmexpress}{1.00}
\newcommand{\weggliUsageRecallOsmexpress}{0.86}
\newcommand{\weggliUsagePrecisionOsmexpress}{0.76}

\newcommand{\weggliUsageRecallKnot}{1.0}
\newcommand{\weggliUsagePrecisionKnot}{0.98}

\newcommand{\grepTotalDistinctAPIKrb}{219}
\newcommand{\weggliTotalDistinctAPIKrb}{228}
\newcommand{\grepTotalDistinctAPIRecorder}{119}
\newcommand{\weggliTotalDistinctAPIRecorder}{106}
\newcommand{\grepTotalDistinctAPIKnot}{149}
\newcommand{\weggliTotalDistinctAPIKnot}{147}

\newcommand{\totalgreplmdbkrb}{69}
\newcommand{\totalwegglilmdbkrb}{69}
\newcommand{\totallibtoollmdbkrb}{69}
\newcommand{\missedgreplmdbkrb}{0}
\newcommand{\missedwegglilmdbkrb}{0}
\newcommand{\ingrepandlibtoollmdbkrb}{69}
\newcommand{\inweggliandlibtoollmdbkrb}{69}
\newcommand{\totalgreplmdbrecorder}{54}
\newcommand{\totalwegglilmdbrecorder}{54}
\newcommand{\totallibtoollmdbrecorder}{54}
\newcommand{\missedgreplmdbrecorder}{0}
\newcommand{\missedwegglilmdbrecorder}{0}
\newcommand{\ingrepandlibtoollmdbrecorder}{54}
\newcommand{\inweggliandlibtoollmdbrecorder}{54}
\newcommand{\totalgreplmdbknot}{72}
\newcommand{\totalwegglilmdbknot}{72}
\newcommand{\totallibtoollmdbknot}{71}
\newcommand{\missedgreplmdbknot}{0}
\newcommand{\missedwegglilmdbknot}{0}
\newcommand{\ingrepandlibtoollmdbknot}{71}
\newcommand{\inweggliandlibtoollmdbknot}{71}
\newcommand{\tbltotalgreplmdbkrb}{20}
\newcommand{\tbltotalwegglilmdbkrb}{20}
\newcommand{\tbltotallibtoollmdbkrb}{20}
\newcommand{\tblingrepandlibtoollmdbkrb}{20}
\newcommand{\tblinweggliandlibtoollmdbkrb}{20}
\newcommand{\tbltotalgreplmdbrecorder}{16}
\newcommand{\tbltotalwegglilmdbrecorder}{16}
\newcommand{\tbltotallibtoollmdbrecorder}{16}
\newcommand{\tblingrepandlibtoollmdbrecorder}{16}
\newcommand{\tblinweggliandlibtoollmdbrecorder}{16}
\newcommand{\tbltotalgreplmdbknot}{24}
\newcommand{\tbltotalwegglilmdbknot}{24}
\newcommand{\tbltotallibtoollmdbliknot}{24}
\newcommand{\tblingrepandlibtoollmdbliknot}{24}
\newcommand{\tblinweggliandlibtoollmdbliknot}{24}

\newcommand{\knotusagepercentage}{43\%\xspace}
\newcommand{\sfmlusagepercentage}{100\%\xspace}

\newcommand{\totalsqliteusages}{949,717\xspace}
\newcommand{\totalluajitusages}{346,784\xspace}
\newcommand{\totaljemallocusages}{331,430\xspace}
\newcommand{\totalncursesusages}{157,993\xspace}
\newcommand{\totalsslusages}{76,833\xspace}
\newcommand{\totalmbedtlsusages}{64,427\xspace}
\newcommand{\totalfftwusages}{735\xspace}
\newcommand{\totalfreetypeusages}{9,076\xspace}
\newcommand{\totalglewusages}{1,626\xspace}
\newcommand{\totalhdfusages}{24,760\xspace}
\newcommand{\totalhidapiusages}{8,378\xspace}
\newcommand{\totallmdbusages}{3,986\xspace}
\newcommand{\totalvorbisusages}{6361\xspace}
\newcommand{\totalxxhashusages}{3132\xspace}
\newcommand{\totalzipusages}{908\xspace}

\title[Understanding API Usage and Testing in C Libraries]{Understanding API Usage and Testing:\\An Empirical Study of C Libraries}

\begin{abstract}
  For library developers, understanding how their Application Programming Interfaces (APIs) are used in the field can be invaluable. 
  Knowing how clients are using their APIs allows for data-driven decisions on prioritising bug reports, feature requests, and testing activities.
  For example, the priority of a bug report concerning an API can be partly determined by how widely that API is used.

In this paper, we present an empirical study in which we analyse API usage across \totalLibs popular open-source C libraries, such as \ssl and \sqlite, with a combined total of \actualNonUniqueClients C/C++ clients.
We compare API usage by clients with how well library test suites exercise the APIs to offer actionable insights for library developers. 

To our knowledge, this is the first study that compares API usage and API testing at scale for the C/C++ ecosystem.
Our study shows that library developers do not prioritise their effort based on how clients use their API, with popular APIs often poorly tested.
For example, in \lmdb, a popular key-value store, \lmdbusedanduncoveredapispercent of the APIs are used by clients but not tested by the library test suite.  
We further show that client test suites can be leveraged to improve library testing---\eg improving coverage in \lmdb by \lmdbimprovedcov---with the important advantage that those tests are representative of how the APIs are used in the field.

For our empirical study, we have developed \libprobe, a framework that can be used to analyse a large corpus of clients for a given library and produce various metrics useful to library developers.
\end{abstract}


\maketitle


\section{Introduction}%
\label{sec:intro}
Libraries provide reusable code for many applications.
As a library becomes more popular, the demands on library developers in terms of fixing bugs, implementing new features, and testing the code increase. 
Understanding how the library's Application Programming Interfaces (APIs) are used can provide invaluable insight for developers.
In particular, library developers would be able to prioritise their time and effort according to data retrieved from a representative sample of clients of the library.
For instance, prioritising bug reports and feature requests is a difficult challenge that has attracted significant research~\cite{taskassessor,codebadsmells,bugreportprioritisation,surveybugpri}.
Without sufficient information about API usage, library developers can spend time and effort maintaining features that are never used by clients.

To better understand how API usage information can help library developers, we have conducted a large-scale empirical study of \totalLibs popular open-source C libraries with a combined total of \actualNonUniqueClients C/C++ clients.
Our empirical study is enabled by \libprobe, a framework we have developed and made available as open source, to analyse library usage information across a large number of clients.

Our study aims to understand how library APIs are used in the field, how well they are tested by the library test suites, and whether there is a correlation between the two.
It also aims to understand whether the size, in terms of lines of code in an API has an impact on API test coverage.
Finally, we investigate whether client test suites could be leveraged to improve testing of API implementations.

In our study, we define an API of a library as an entry function (exported symbol) of that library. 
We measure the size and coverage of an API implementation by considering only the code within the entry function itself, excluding any code in its callees. 
This design choice is further discussed in \S\ref{sec:lib-proc}.
For succinctness, in the rest of the paper, we use the terms \emph{API implementation}, \emph{API size} and \emph{API coverage} to refer, respectively, to the code implementation, number of lines of code, and percentage of lines of code covered in the entry function.
\subsection{Research Questions}
\label{sec:rqs}

Our empirical study answers the following research questions:

\newcommand{\rqOne}{RQ1: What is the distribution of library API uses across clients?}
\mypara{\rqOne}
What percentage of a library's APIs are used by clients and how commonly do clients use the full set of APIs from that library?
Does API utilisation depend on the number of APIs offered by the library?
Is there a large difference in number of uses between the most and least used APIs?

\newcommand{\rqThree}{RQ2: How well are API implementations tested by the library test suite? Does API implementation size matter?}
\mypara{\rqThree}
Is there a correlation between the API implementation size and the API test coverage achieved by the library test suite?

\newcommand{\rqFour}{RQ3: Are APIs widely used by clients also well tested?}
\mypara{\rqFour}
Is there a correlation between the number of clients using an API and the API test coverage achieved by the library test suite?

\newcommand{\rqFive}{RQ4: Can API coverage be improved by using the client test suites?}
\mypara{\rqFive}
Is it possible to leverage the client test suites to better test libraries?
This would be particularly valuable as the client test suites are more likely to reflect how the libraries are used in practice.

\subsection{Contributions}
To our knowledge, this is the first large-scale empirical study for C/C++ that tries to understand how library APIs are used in practice and correlate that information with test coverage, based on a large number of client applications.
Our study includes \totalLibs libraries and \actualNonUniqueClients clients hosted on the popular GitHub platform.
Our major findings can be summarised as follows:
\begin{enumerate}[leftmargin=*]
\item \textit{API testing compared to usage:} Many libraries have APIs that are used by a large number of clients, yet the testing of those API implementations is poor.
  Conversely, there are many APIs with few or no clients which are well-tested by the library developers.
  This shows that library developers can benefit from metrics about API usage to prioritise their testing efforts.

\item \textit{API utilisation by clients:} Most libraries have unused APIs, and the percentage of unused APIs does not depend on the number of APIs offered.
  This information can be used to improve API design and retire unnecessary APIs.
  
\item \textit{Improving API coverage using client test suites:} The test suites of library clients can be leveraged to improve API coverage.
  Such tests have the further advantage of being representative of how the APIs are used in the field. 
  For example, we could improve coverage in \vorbis by \vorbisimprovedcov reaching \vorbisnewlycoveredapis previously untested APIs by leveraging the test suite of a single client.
\end{enumerate}

\vspace{0.01in}
To conduct our study, we have developed a fast, lightweight framework for large-scale API usage analysis of C libraries and C/C++ clients that produces helpful metrics for library developers.
We make our framework, \libprobe, and the results of our empirical study available as an artifact~\cite{artifact}.


\section{LibProbe}
\label{sec:design}
\begin{figure*}[t]
    \centering
    \includegraphics[width=\linewidth]{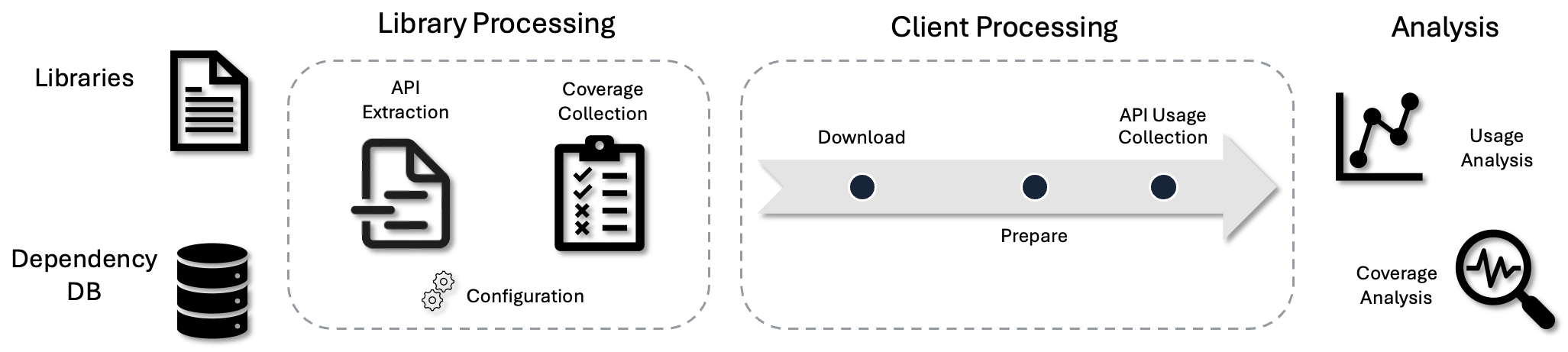}
    \caption{\libprobe architecture.}
    \label{fig:architecture}
\end{figure*}

The high-level architecture of \libprobe is shown in Figure~\ref{fig:architecture}.
The inputs to \libprobe are the libraries of interest and a dependency database.
The latter consists of $(C, L)$ pairs, signifying that client $C$ uses library $L$, where $C$ can itself be a library.
We use the dependency database provided by \ccscanner~\cite{ccscanner}, which includes \github repositories for all entries. 
We discuss this database further in \S\ref{sec:methodology}.

\libprobe starts by processing each library to obtain its set of APIs and the test coverage achieved on the code of each API by the library's test suite (\S\ref{sec:lib-proc}).
It then uses information from the dependency database to download and prepare for analysis all the known clients of the library (\S\ref{sec:client-prep}).
\libprobe then analyses each client to extract all the uses of the library APIs (\S\ref{sec:usages}).
Optionally, \libprobe determines the extra coverage achieved by the client in the library under analysis.
Finally, \libprobe processes and summarises the collected data in the form of \textit{JSON} files and graphs.

\mypara{Scope and requirements.}
\libprobe is meant to analyse popular libraries, which may have dozens if not hundreds of C/C++ clients.
This imposes several constraints on its design.

First and foremost, we need to avoid the need to build each client, which is both error-prone and expensive.
This means we cannot use compiler infrastructures to parse and analyse the code.
Instead, we rely on simple lexical analysis of the code to collect usage information.
This is one of the reasons for which we restrict \libprobe to C libraries, as a lexical search for C++ APIs is more error-prone.
However, \libprobe can process both C and C++ clients. 
We discuss this aspect in more detail in \S\ref{sec:usages}.

Second, we need a way to distinguish between client and library code in the common scenario in which the client codebase incorporates the library code.
We discuss this aspect in \S\ref{sec:client-prep}.

Third, we need to be able to build the library itself into a shared library, as \libprobe uses the exported symbols from shared library archives as a step to determine the set of APIs of a library. 
As such, header-only libraries cannot be processed by \libprobe.

\subsection{Library Processing}
\label{sec:lib-proc}

The first stage of \libprobe is to extract the set of exported symbols from each shared library file\footnote{We do so by running \texttt{nm -CD library.so | grep " T "} on the shared library.}, which is often a superset of the APIs documented for that library.
This is because some symbols are used only for communication between different modules of a library.
For instance, this was the case for \ncurses~\cite{ncurses-site} and \mbedtls~\cite{mbedtls-site}, as documented in our conversations with their developers~\cite{ncruses-discussion, mbedtls-discussion}.
In the case of \mbedtls, the developers mentioned that functions exported and not part of the API are not made private or have no name mangling because they never got around to doing this in their build scripts.
To tackle this, we filter out symbols that are not part of the library API by excluding those that do not appear in the headers of the library after installation.
We do not solely rely on the installed header files since that would be less precise in the presence of macros. 


\libprobe stores all APIs in a database to search for them later in the clients.
It will also record the size (in number of ELOC) and test coverage (if provided) of each API implementation, by processing all the coverage files present in the library's directory.
In our study, we measure test coverage in terms of \emph{line coverage}, as reported by \gcov~\cite{gcov} and \lcov~\cite{lcov}.
While high line coverage alone is not sufficient, it is nevertheless necessary; library test suites cannot find issues in code that is not exercised. 
It is worth noting that we process all \textit{.gcno} files which are generated when the library is compiled. 
This can include test or other auxiliary files so our coverage reporting is an overapproximation of the actual library coverage. 
Generally, we make a best effort to identify specific directories that are part of the core library source code, otherwise, we process all compiled files and report the total coverage on those files.

The entry functions of an API represent a critical interface between the library and its clients.
Such functions often perform input validation and high-level decisions concerning the functionality offered by the API. 
Depending on how the code is structured, in some cases the entry function may not reflect the complexity of the API implementation (because most of the core logic may be in the callees).
We explored measuring the full size of an API implementation by aggregating the number of ELOC of all callees used by the API recursively. 
That approach suffered from two fundamental problems. 
First, callees can be shared between different APIs, which would make calculating the coverage for each API inaccurate.
For instance, consider an API which is never tested, but for which its callees are exercised through other APIs; the API coverage in this case would be considered non-zero, when in fact the API is not exercised at all by the test suite.
This problem extends recursively, in the same way, to the callees themselves. 
Second, the size of the API, in ELOC, would often be a large overestimate rather than an accurate measurement, as only a small part of the code of the callees might be used by the given API .
In summary, we believe it is more meaningful to restrict our code coverage and size measurement to the API entry functions.

\subsection{Client Preparation}
\label{sec:client-prep}

Clients of a library sometimes include the library code in their own codebase to make it easier to build the code, and/or to ensure the use of specific library versions.
To be able to get an accurate assessment of API usage, we have to exclude client directories which contain the library code.
To do this, we use two approaches. 

Our first approach handles Git submodules~\cite{git-website}, when they are present in the client's codebase.
We read the \textit{.gitsubmodules} file and exclude all paths in it. 
This could include other dependencies, which is fine, as we are only interested in the client code itself. 

Our second approach handles the case where the library code is added as a sub-directory. 
We list all the directories in the library's repository which include source files and collect all the file names in those directories recursively.
Following that, we look for matches in the client's code.
Given a client directory, we exclude it if 80\% of the files in the client directory exist in the library directory (as long as the latter has more than two files).
The reason for allowing a partial match is that sometimes clients use an old version of a library that may contain slightly different files. 
We arrived to the threshold of 80\% through several experiments across library-client pairs to maximise accuracy. 

Some libraries have a small number of source files with distinctive names, in which case we explicitly exclude them using their name.
The reason we do not do this for all libraries is that many files have generic names, with different files being given the same name in the library and the client.

\subsection{API Usage Collection}
\label{sec:usages}

Our empirical study involves a large number of clients---\ccscanner reports \totalClients different clients for our libraries. 
Therefore, our API usage collection method needs to be both fast and lightweight.
In particular, this excludes techniques which rely on building each client codebase, such as those using a compiler framework like \clang~\cite{clang}.
In addition to taking considerable time, building such a large number of clients would likely be infeasible given the variety of build systems and dependencies involved. 

Therefore, we set as a strict requirement to use a simple lightweight analysis based on textual search.
We considered two approaches: one based on the \grep~\cite{grep-website} text search utility, and the other based on the \weggli~\cite{weggli-website} semantic search tool.

\lstset{
    basicstyle=\ttfamily\small,
    columns=flexible,
    numbers=none,
    breaklines=true
}

\mypara{\grep search.}
For each library API, we use a multi-step \grep pipeline on all source files in each client directory.
We start by excluding comments by running the following command on the client root directory:
\begin{lstlisting}[basicstyle=\small]
grep -rEIv \/\/[^\n]*|\/\*.*[\*\/]?|^\s\* --include=*.cc --include=*.c --include=*.cpp --include=*.cxx --include=*.hh --include=*.h --include=*.hpp --include=*hxx --exclude=<client_dir> ..
\end{lstlisting}

This command outputs text in source files that excludes comments and specific client directories that we identified in \S\ref{sec:client-prep}.
We then search for uses of an API across this text, while excluding string literal matches by the running the following two commands:
\begin{lstlisting}[basicstyle=\small, frame=none]
  grep -E \b<api>\s?\(
  grep -Ev ".*<api>.*"
  \end{lstlisting}




\mypara{\weggli search.}
\weggli~\cite{weggli-website} is a semantic search tool for C and C++ codebases, which is built on top of \texttt{tree-sitter}~\cite{treesitter-website}.
We used \weggli to search for each library API in each client source file, excluding again the directories that were identified during the client preparation stage.
   
We compared the \grep and \weggli approaches to each other and to a tool built on top of \libtool~\cite{libtooling}.
As discussed earlier, the latter would be too expensive to apply in practice but is used as a comparison baseline.
Our \libtool-based tool, further referred to as \textit{libtool}, fetches all call expressions invoking APIs from the library of interest.

We randomly selected two libraries, each with five clients, to compare the three approaches.
We selected: \mbedtls~\cite{mbedtls-site} with clients \uacme~\cite{uacme-site}, \curl~\cite{curl-site}, \openvpn~\cite{openvpn-site}, \lighttpd~\cite{lighttpd-site} and \libcoap~\cite{libcoap-site}; and \lmdb~\cite{lmdb-site} with clients \krb~\cite{krb5-site}, \recorder~\cite{recorder-site}, \libetpan~\cite{libetpan-site}, \fapolicyd~\cite{fapolicyd-site} and \osmexpress~\cite{osmexpress-site}.

We performed two types of analysis; one on identifying the distinct APIs used, and the other on counting the number of uses for each API.
The former is useful for determining how many of the library's APIs are being used by a client, while the latter for understanding the popularity of each API.
We calculate Precision and Recall as
\[
\textit{P}_{\scriptscriptstyle D/T} = \frac{\textit{Tp}_{\scriptscriptstyle D/T}}{\textit{Tp}_{\scriptscriptstyle D/T} + \textit{Fp}_{\scriptscriptstyle D/T}}
\quad \quad
\textit{R}_{\scriptscriptstyle D/T} = \frac{\textit{Tp}_{\scriptscriptstyle D/T}}{\textit{Tp}_{\scriptscriptstyle D/T} + \textit{Fn}_{\scriptscriptstyle D/T}}
\]

\textit{P}$_{\scriptscriptstyle D}$ and \textit{R}$_{\scriptscriptstyle D}$ represents precision and recall for distinct API identification while \textit{P}$_{\scriptscriptstyle T}$ and \textit{R}$_{\scriptscriptstyle T}$ represents precision and recall for total uses for each API identified.
We define \textit{Tp}$_{\scriptscriptstyle D}$ as the the number of distinct APIs identified by both the tool (\grep or \weggli) and \textit{libtool}; and \textit{Fp}$_{\scriptscriptstyle D}$ as the number of distinct APIs identified by the tool but not by \textit{libtool}.
\textit{Fn}$_{\scriptscriptstyle D}$ is defined as the number of distinct APIs identified by \textit{libtool} but not identified by the tool.
\textit{Tp}$_{\scriptscriptstyle T}$, \textit{Fp}$_{\scriptscriptstyle T}$, and \textit{Fn}$_{\scriptscriptstyle T}$ are defined similarly for total API uses.


\begin{table}[t]
  \centering
  \caption{\grep and \weggli precision and recall on clients of \mbedtls relative to \textit{libtool}.}
  \resizebox{\columnwidth}{!}{ 
   \begin{tabular}{l c c | c c}
     \toprule
     & \multicolumn{2}{c}{Distinct API Uses} & \multicolumn{2}{c}{Total API Uses} \\ 
     \cmidrule(lr){2-3} \cmidrule(lr){4-5}
     & \grep & \weggli & \grep & \weggli \\ 
      Client & \textit{P}$_{\scriptscriptstyle D}$ / \textit{R}$_{\scriptscriptstyle D}$ & \textit{P}$_{\scriptscriptstyle D}$ / \textit{R}$_{\scriptscriptstyle D}$ & \textit{P}$_{\scriptscriptstyle T}$ / \textit{R}$_{\scriptscriptstyle T}$ & \textit{P}$_{\scriptscriptstyle T}$ / \textit{R}$_{\scriptscriptstyle T}$ \\
      \midrule
    \uacme& \grepApiPrecisionUacme{} / \grepApiRecallUacme & \weggliApiPrecisionUacme{} / \weggliApiRecallUacme & \grepUsagePrecisionUacme{} / \grepUsageRecallUacme & \weggliUsagePrecisionUacme{} / \weggliUsageRecallUacme \\
    \curl& \grepApiPrecisionCurl{} / \grepApiRecallCurl & \weggliApiPrecisionCurl{} / \weggliApiRecallCurl & \grepUsagePrecisionCurl{} / \grepUsageRecallCurl & \weggliUsagePrecisionCurl{} / \weggliUsageRecallCurl \\
    \openvpn& \grepApiPrecisionOpenvpn{} / \grepApiRecallOpenvpn & \weggliApiPrecisionOpenvpn{} / \weggliApiRecallOpenvpn & \grepUsagePrecisionOpenvpn{} / \grepUsageRecallOpenvpn & \weggliUsagePrecisionOpenvpn{} / \weggliUsageRecallOpenvpn \\
     \lighttpd& \grepApiPrecisionLighttpd{} / \grepApiRecallLighttpd & \weggliApiPrecisionLighttpd{} / \weggliApiRecallLighttpd & \grepUsagePrecisionLighttpd{} / \grepUsageRecallLighttpd & \weggliUsagePrecisionLighttpd{} / \weggliUsageRecallLighttpd \\
    \libcoap& \grepApiPrecisionLibcoap{} / \grepApiRecallLibcoap & \weggliApiPrecisionLibcoap{} / \weggliApiRecallLibcoap & \grepUsagePrecisionLibcoap{} / \grepUsageRecallLibcoap & \weggliUsagePrecisionLibcoap{} / \weggliUsageRecallLibcoap \\  
    \bottomrule
  \end{tabular}
  }
  \label{tbl:mbedtlseval}
\end{table}

 \begin{table}[t]
  \centering
  \caption{\grep and \weggli precision and recall on clients of \lmdb relative to \textit{libtool}.}
  \resizebox{\columnwidth}{!}{ 
  \begin{tabular}{l c c | c c}
    \toprule
    & \multicolumn{2}{c}{Distinct API Uses} & \multicolumn{2}{c}{Total API Uses} \\ 
    \cmidrule(lr){2-3} \cmidrule(lr){4-5}
    & \grep & \weggli & \grep & \weggli \\ 
     Client & \textit{P}$_{\scriptscriptstyle D}$/ \textit{R}$_{\scriptscriptstyle D}$ & \textit{P}$_{\scriptscriptstyle D}$/ \textit{R}$_{\scriptscriptstyle D}$ & \textit{P}$_{\scriptscriptstyle T}$/ \textit{R}$_{\scriptscriptstyle T}$ & \textit{P}$_{\scriptscriptstyle T}$/ \textit{R}$_{\scriptscriptstyle T}$  \\
     \midrule
    \krb& \grepApiPrecisionKrb{} / \grepApiRecallKrb & \weggliApiPrecisionKrb{} / \weggliApiRecallKrb & \grepUsagePrecisionKrb{} / \grepUsageRecallKrb & \weggliUsagePrecisionKrb{} / \weggliUsageRecallKrb\\
    \recorder& \grepApiPrecisionRecorder{} / \grepApiRecallRecorder & \weggliApiPrecisionRecorder{} / \weggliApiRecallRecorder & \grepUsagePrecisionRecorder{} / \grepUsageRecallRecorder & \weggliUsagePrecisionRecorder{} / \weggliUsageRecallRecorder\\
    \libetpan& \grepApiPrecisionLibetpan{} / \grepApiRecallLibetpan & \weggliApiPrecisionLibetpan{} / \weggliApiRecallLibetpan & \grepUsagePrecisionLibetpan{} / \grepUsageRecallLibetpan & \weggliUsagePrecisionLibetpan{} / \weggliUsageRecallLibetpan\\
    \fapolicyd& \grepApiPrecisionFapolicyd{} / \grepApiRecallFapolicyd & \weggliApiPrecisionFapolicyd{} / \weggliApiRecallFapolicyd & \grepUsagePrecisionFapolicyd{} / \grepUsageRecallFapolicyd & \weggliUsagePrecisionFapolicyd{} / \weggliUsageRecallFapolicyd\\
    \osmexpress& \grepApiPrecisionOsmexpress{} / \grepApiRecallOsmexpress & \weggliApiPrecisionOsmexpress{} / \weggliApiRecallOsmexpress & \grepUsagePrecisionOsmexpress{} / \grepUsageRecallOsmexpress & \weggliUsagePrecisionOsmexpress{} / \weggliUsageRecallOsmexpress\\
   \bottomrule
  \end{tabular}
   }
  \label{tbl:lmdbeval}
 \end{table}

 Tables~\ref{tbl:mbedtlseval} and \ref{tbl:lmdbeval} show the results.
 \grep generally performed better than \weggli for all clients of \mbedtls except for \lighttpd, where both tools had a low precision.
For clients of \lmdb, Table~\ref{tbl:lmdbeval} shows that both tools were largely similar in performance except for \libetpan and \osmexpress.

 Looking closer at the low precision of both tools on \lighttpd we found that \grep reported 37 false positive distinct APIs while \weggli reported 39.
Almost all of the reported APIs were either inside \texttt{\#ifdef} directives that look for a certain \mbedtls version/configuration or were in source files that were not part of the build of the client.
Since \textit{libtool} runs on the source \textit{after} it gets pre-processed and takes into consideration the build configuration, some code gets removed.
As such, if an API is within an \texttt{\#ifdef} that looks for a certain library version or perhaps a debug build, \textit{libtool} will miss such uses.
This was also reflected when looking at the total uses reported by both \grep and \weggli.
\grep reported 66 false positive uses compared to \textit{libtool} while \weggli reported 72.
The majority of the false positives were due to either \texttt{\#ifdef} directives looking for a version of \mbedtls or due to files not included in the build. 
It is possible to argue that these uses are not really false positives. 
The uses are in client's source files but not included in the standard build configuration or only used when certain conditions are met.
This does still mean that under certain conditions such APIs could be used by the client. 

\weggli performed better than \grep on \lighttpd, \libetpan and \fapolicyd.
These differences were due to two main factors.
First, \grep discards all the lines with comments. 
Therefore, if a comment is on the same line as a call, it incorrectly gets discarded.
While we could improve this aspect, it is difficult to come up with a general solution without parsing the code. 
The second source of false negatives is when an API function is passed as a function pointer. 
This is due to the regular expression we used to find API uses, which only looks for call expressions; 
through experimentation, we found that removing this restriction resulted in a higher number of false positives.

In the majority of cases, \grep performed better than \weggli.
For instance, in \libcoap, \weggli was unable to identify the usage of 4 distinct APIs (false negatives) while \grep reported all APIs used by the \textit{libtool}.
Similarly, in term of total uses, \weggli missed 12 uses, which \grep successfully reported.

Analysing \weggli's misses, we identified two causes. 
A large fraction of the misses come from \weggli's failure to parse some functions that use \texttt{\#ifdef} directives heavily.
Usage of \texttt{\#ifdef} directives inside functions resulted in failures of \texttt{tree-sitter}, which \weggli uses to generate an AST. 
Since \grep is lexical, it has no issues identifying these uses.
Another limitation of \weggli is with respect to macros. 
\weggli is unable to process macro definitions, and as such misses completely API uses that are wrapped in a macro.
We confirmed this by raising an issue on the \weggli \github project~\cite{weggli-issue}.

In summary, both \weggli and \grep have some limitations, but overall we were more concerned about \weggli's inability to process macros and \texttt{\#ifdef} directives, which are common in C code.
Combined with the fact that \weggli does not seem to be actively developed anymore (which means that any issues encountered during our study could be difficult to resolve), we have decided to use \grep in \libprobe.

\section{Empirical Study}
\label{sec:eval}
In this section, we present the results of our empirical study involving \totalLibs libraries and \actualNonUniqueClients clients.
We start by presenting our methodology in \S\ref{sec:methodology}, after which we present the results for the first research question in \S\ref{sec:api_usage} and for the last three in \S\ref{sec:coverage}.

\subsection{Methodology}
\label{sec:methodology}

\mypara{Dependency database.}
In our study, we make use of \ccscanner~\cite{ccscanner}, which provides a database of dependencies for 24K C/C++ \github projects.
In particular, each entry in the \ccscanner database consists of a \github repository and its dependencies.





\begin{figure}[t]
  \centering
  \includegraphics[width=\columnwidth]{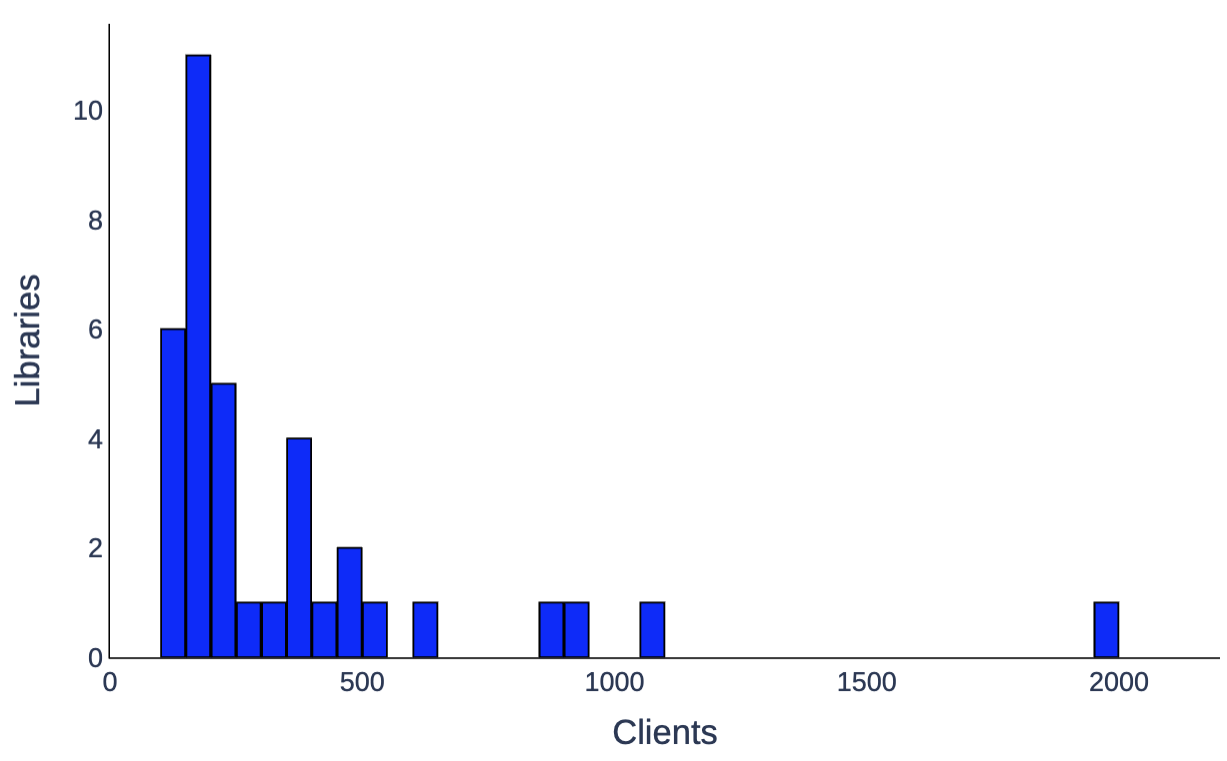}
  \caption{Histogram of the number of libraries from \ccscanner with at least 100 clients.}
  \label{fig:usagedistrib}
\end{figure}

\mypara{Library selection.}
%
From the total of 24K repositories, 229 were not available anymore on \github, and thus we discarded them.
We then identified the number of C repositories in the \ccscanner dataset.
Using each repository's \github metadata, we identified 10,291 repositories that had C as their main language.
Out of these, 2,520 were dependencies of repositories in the \ccscanner dataset. 
%
In terms of number of clients, 2,067 (82\%) of the dependencies had less than 10 clients, 354 (14\%) had between 10 to 49 clients, 54 (2\%) between 50 and 99 clients, and 45 (2\%) at least 100 clients. 
In our study, we are interested in popular libraries, so we have chosen to focus on the 45 libraries with at least 100 clients.

We manually reviewed the 45 dependencies, and eliminated those which are clearly not libraries, such as \cmake, and \systemd.
In total, we identified 8 such non-libraries, leaving us with 37 libraries with at least 100 clients.

Figure~\ref{fig:usagedistrib} shows a histogram of the number of libraries with at least 100 clients. 
To avoid long processing cycles in our study, we select the 32 libraries that have between 100 and 550 clients, which represent the majority of the libraries (the contiguous bars on the left in Figure~\ref{fig:usagedistrib}).

In order to identify the APIs available for use by clients, \libprobe requires each library repository to be built into a shared library and installed.
We manually attempted to build each of the 32 libraries. 
However, we could not build 11 of them, as some libraries were header-only, while others required specific system-level environment configurations, such as the Android SDK. 

This process left us with \totalLibs libraries to use for our study, which are shown in Table~\ref{tbl:alllibs}, together with their size in ELOC, and the number of APIs they provide.
Our selection includes libraries with diverse applications such as encryption (\ssl~\cite{ssl-site}), database management (\sqlite~\cite{sqlite-site}), media (\sdl~\cite{sdl-site}), compression (\zip~\cite{zip-site}), rendering (\freetype~\cite{freetype-site}) and general purpose libraries (\glib~\cite{glib-site}). 
The ELOC measurements are provided by \gcov and \lcov, so they reflect the size of these libraries as compiled on our platform. 


\begin{table}[t]
  \centering
  \caption{Libraries analysed, together with their number of APIs, number of clients reported by \ccscanner, and some examples of clients.
  }
  \begin{tabularx}{\linewidth}{l r r l}
    \toprule
    Library & APIs & Clients & Example clients\\
    \midrule
    \ffmpeg & \ffmpegapisize & \ffmpegclients & RoboMaster, Telegram\\
    \fftw & \fftwapisize & \fftwclients & Atomify, Synfig, Cava\\
    \freetype & \freetypeapisize & \freetypeclients & Ogre, OpenHarmony\\
    \glew & \glewapisize & \glewclients & openglText, imgui\\
    \glib & \glibapisize & \glibclients & Mutter, GTK\\
    \gsl & \gslapisize & \gslclients & OpenPilot, GDL\\
    \hdf & \hdfapisize & \hdfclients & RDPFuzz, OpenCV\\
    \hidapi & \hidapiapisize & \hidapiclients & OpenSCAD, OpenFPGA\\
    \jemalloc & \jemallocapisize & \jemallocclients & SpiderMonkey\\
    \lmdb & \lmdbapisize & \lmdbclients & PowerDNS, Caffe, Dali\\
    \luajit & \luajitapisize & \luajitclients & Redis-storage\\
    \lz & \lzapisize & \lzclients & FreeBSD, NVBio\\
    \mbedtls & \mbedtlsapisize & \mbedtlsclients & OpenVPN3, Mqtt-c\\
    \ncurses & \ncursesapisize & \ncursesclients & WeeChat, Heimdal\\
    \ssl & \sslapisize & \sslclients & Telegram iOS, Moai\\
    \sdl & \sdlapisize & \sdlclients & AliOS, DreamShell\\
    \sqlite & \sqliteapisize & \sqliteclients & Gideros, OrangeC\\
    \vorbis & \vorbisapisize & \vorbisclients & Spring, Pindrop\\
    \xxhash & \xxhashapisize & \xxhashclients & FreeBSD, Orbit\\
    \zip & \zipapisize & \zipclients & Assimp, Radare\\
    \zstd & \zstdapisize & \zstdclients & Grok, Qemu\\
    \bottomrule
  \end{tabularx}
  \label{tbl:alllibs}
\end{table}

\vspace{0.07in}
\mypara{Client selection.} For each of the \totalLibs libraries, we select all their available clients in the \ccscanner database.
Note that while the libraries are all written in C, the clients can be either C or C++ projects.
Table~\ref{tbl:alllibs} shows the number of clients downloaded and processed for each library and, between parenthesis, the number of clients where we have actually identified API uses.

In total, \ccscanner listed \totalClients unique repositories as clients of our \totalLibs libraries.
As seen in Table~\ref{tbl:alllibs}, there is a substantial difference between the number of clients reported by \ccscanner and those confirmed by us for each library.
We identified two reasons for this discrepancy.
First, we noticed that some of the dependency information in \ccscanner is incorrect.
This could be due to the evolution of the clients over time since the publication of the \ccscanner dataset.
Second, we explicitly remove third-party dependencies added via Git submodules, as we are interested only in how the client itself uses libraries, while \ccscanner would consider that the client uses a library even when only one of its third-party dependencies does. 

Even after filtering out these clients, we are left with \actualNonUniqueClients clients across all libraries, of which \actualUniqueClients are unique.

\mypara{Running \libprobe.} For each library and client, we run the three \libprobe stages discussed earlier: library preparation (\S\ref{sec:lib-proc}), client preparation (\S\ref{sec:client-prep}) and API usage collection (\S\ref{sec:usages}).
Running \libprobe took approximately 96 hours on a machine with an AMD EPYC 7302P 4 Core Processor at 3.6GHz, and 32GB RAM.
The OS was Ubuntu 22.04 LTS, and our compiler was \textsc{GCC} 11.4.0.

\subsection{Usage Analysis}

To address the first RQ, we study how APIs are used in practice, in particular what percentage of a library's APIs are used per client, if API utilisation depends on the number of APIs offered by the library, and if there is a large difference in number of uses between the most and least used APIs.

\label{sec:api_usage}

\begin{figure}[tb]
  \centering
  \includegraphics[width=\columnwidth]{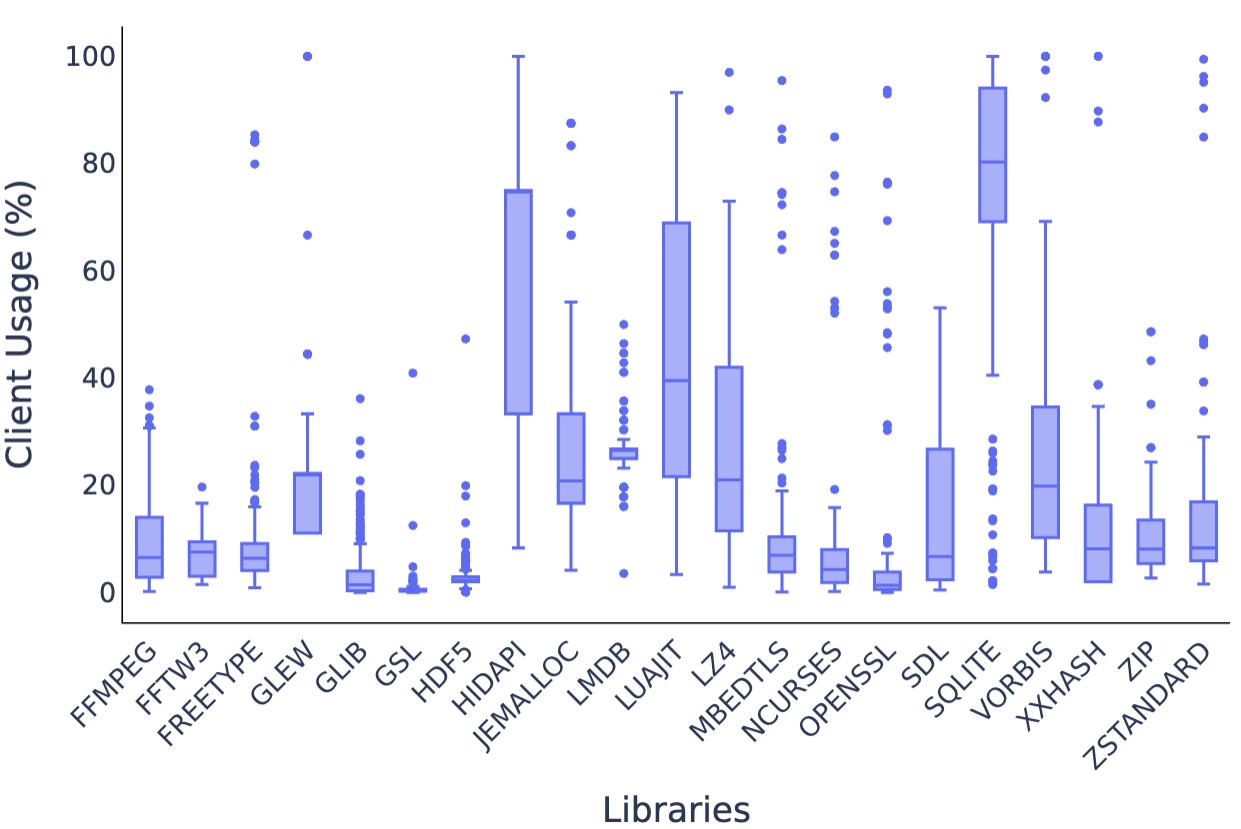}
  \caption{Percentage of library APIs used by each client.}
  \label{fig:clientusages}
\end{figure}

\begin{figure}[tb]
  \centering
    \includegraphics[width=\columnwidth]{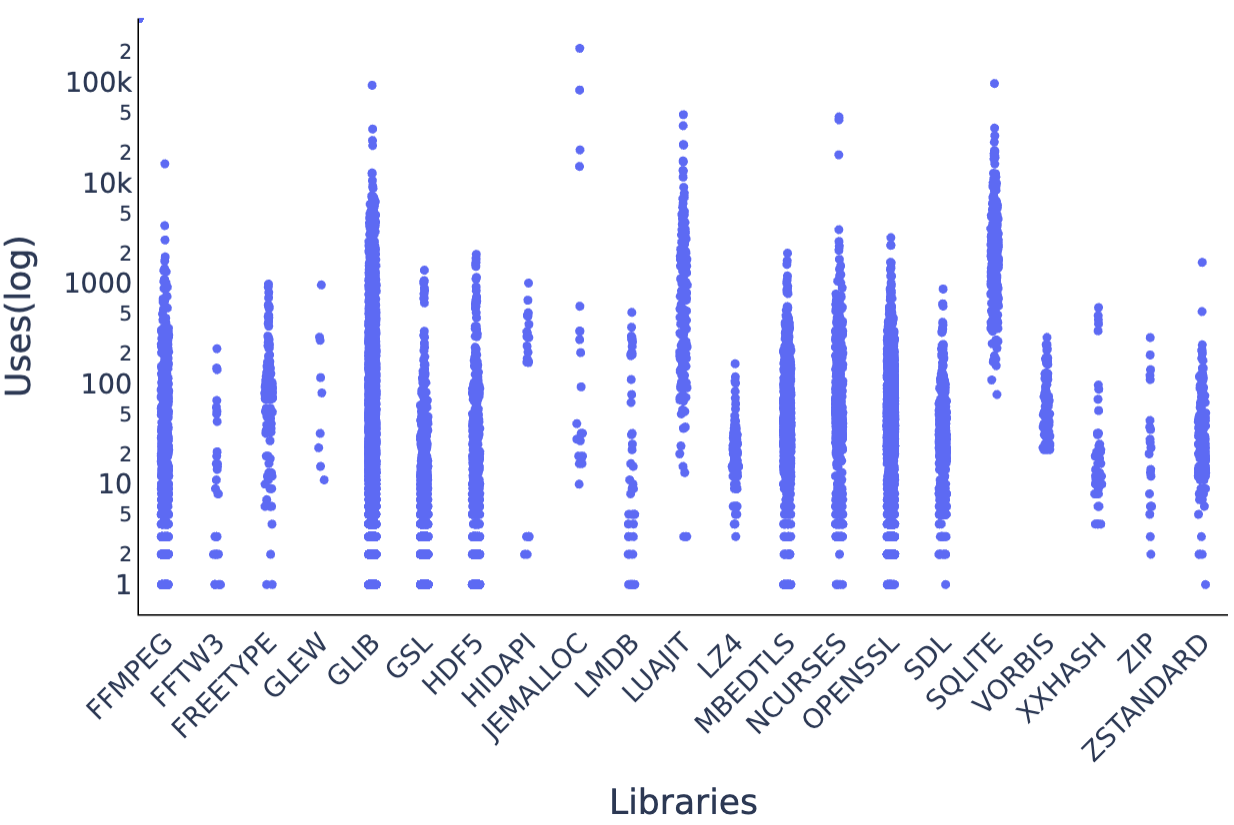}
    \caption{Number of uses for each library API (log scale).}
  \label{fig:invocations}
\end{figure}

\mypara{\rqOne}
\label{subsec:client_usage_analysis}
\label{sec:popular_apis}
%

\mypara{Percentage of APIs used per client.}
Figure~\ref{fig:clientusages} shows the percentage of library APIs used by \textit{each} client.
A few libraries, like \xxhash~\cite{xxhash-site}, \hidapi~\cite{hidapi-site}, \vorbis~\cite{vorbis-site}, \sqlite~\cite{sqlite-site}, \glew~\cite{glew-site} and \zstd~\cite{zstd-site}, had clients that used 100\% of their APIs. 
All of those libraries have a relatively small number of APIs, as shown in Table~\ref{tbl:alllibs}, except \sqlite which has \sqliteapisize APIs.
For most libraries, the upper quartile usage is under 40\% of the library API.
Some libraries, such as \fftw~\cite{fftw-site}, \gsl~\cite{gsl-site} and \glib~\cite{glib-site}, do not have clients that exceed 40\% API utilisation.

\begin{table}[tb]
  \centering
  \caption{Unused APIs per library, sorted in descending order of the percentage of unused APIs.}
  \begin{tabular}{l r r r}
    \toprule
    \multirow{2}{*}{Library} &\multicolumn{3}{c}{APIs} \\
                          & Total & Unused  & Unused \% \\
    \midrule
    \hdf & \hdfapisize & \unusedApiHdf & \hdfApiUnusedPercentage \\
    \gsl & \gslapisize & \unusedApiGsl & \gslApiUnusedPercentage \\
    \sdl & \sdlapisize & \unusedApiSdl & \sdlApiUnusedPercentage \\
    \fftw & \fftwapisize & \unusedApiFftw & \fftwApiUnusedPercentage \\
    \zip & \zipapisize & \unusedApiZip & \zipApiUnusedPercentage \\
    \glib & \glibapisize & \unusedApiGlib & \glibApiUnusedPercentage \\
    \ffmpeg & \ffmpegapisize & \unusedApiFFmpeg & \ffmpegApiUnusedPercentage \\
    \lmdb & \lmdbapisize & \unusedApiLmdb & \lmdbApiUnusedPercentage \\
    \lz & \lzapisize & \unusedApiLz & \lzApiUnusedPercentage \\
    \jemalloc & \jemallocapisize & \unusedApiJemalloc & \jemallocApiUnusedPercentage \\
    \mbedtls & \mbedtlsapisize & \unusedApiMbedtls & \mbedtlsApiUnusedPercentage \\
    \freetype & \freetypeapisize & \unusedApiFreetype & \freetypeApiUnusedPercentage \\
    \ssl & \sslapisize & \unusedApiSsl & \sslApiUnusedPercentage \\
    \ncurses & \ncursesapisize & \unusedApiNcurses & \ncursesApiUnusedPercentage \\
    \luajit & \luajitapisize & \unusedApiLuajit & \luajitApiUnusedPercentage \\
    \zstd & \zstdapisize & \unusedApiZstd & \zstdApiUnusedPercentage \\
    \glew & \glewapisize & \unusedApiGlew & \glewApiUnusedPercentage \\
    \hidapi & \hidapiapisize & \unusedApiHidapi & \hidapiApiUnusedPercentage \\
    \sqlite & \sqliteapisize & \unusedApiSqlite & \sqliteApiUnusedPercentage \\
    \vorbis & \vorbisapisize & \unusedApiVorbis & \vorbisApiUnusedPercentage\\
    \xxhash & \xxhashapisize & \unusedApiXxhash & \xxhashApiUnusedPercentage \\
    \bottomrule
  \end{tabular}
  \label{tbl:unusedapis}
\end{table}

\mypara{Unused APIs.}
Table~\ref{tbl:unusedapis} shows the number of unused APIs and the percentage of unused APIs for each library.
Looking at libraries with a large number of APIs, some have very high utilisation, such as \ssl, where only \sslApiUnusedPercentage of its \sslapisize APIs are unused, while others have low utilisation, such as \hdf, where \hdfApiUnusedPercentage of its \hdfapisize APIs are unused.
The same variation holds for libraries with a small number of APIs: for instance, all of \xxhash's \xxhashapisize APIs are used, while \zipApiUnusedPercentage of \zip's \zipapisize APIs are unused.
Clearly, the number of APIs available from a library does not correlate with usage.
This can be due to some libraries requiring clients to use many APIs to achieve a single functionality. 
As can be seen, \libsWithUnusedApis of the \totalLibs libraries have at least one unused API, with the percentage of unused APIs reaching \hdfApiUnusedPercentage for \hdf~\cite{hdf-site} and \gslApiUnusedPercentage for \gsl~\cite{gsl-site}.

We took a closer look at the unused APIs across libraries and found that they fell into three categories:
APIs covering secondary functionality, such as helper and debugging APIs; APIs with a similar functionality to more popular APIs; and less-used modules. 

Libraries like \ssl, \zip, \lmdb and \freetype have many helper-type APIs that were unused.
In the case of \freetype and \lmdb, many seem to be getter- and setter-type functions such as \textit{FT\_Set\_Log\_Handler} and \textit{mdb\_env\_set\_flags}.

Some libraries have several APIs performing more or less the same functionality as other more popular APIs.
For example, users of \zip can use \textit{zip\_close} instead of the unused APIs \textit{zip\_stream\_close} and \textit{zip\_cstream\_close}.

Libraries that offer multiple modules see little usage of the APIs in some of the modules; this is the case in \ssl, \fftw, \hdf and \mbedtls.
For instance, in the case of \fftw, the majority of the unused APIs are related to the \textit{Guru} module.
This seems to offer a fragile API, which users avoid. 
According to the documentation, \textit{``For those users who require the flexibility of the guru interface, it is important that they pay special attention to the documentation lest they shoot themselves in the foot''}~\cite{fftw-guru}.
A simple internet search returns multiple questions on \textit{Stack Overflow} on how to use this interface~\cite{fftw-guru-q1,fftw-guru-q2,fftw-guru-q3}.

\mypara{Number of API uses.}
Figure~\ref{fig:invocations} shows, for each library, the distribution of API uses across clients, for all APIs that are used.
The y-axis is shown in log scale.
We can see that for some libraries, such as \glib~\cite{glib-site}, \ncurses~\cite{ncurses-site} and \jemalloc~\cite{jemalloc-site}, there is a wide difference between the largest and smallest number of uses.

\vspace{0.08in}

The information collected as part of this RQ can be of great value to developers in prioritising API development and testing.
Information about the API utilisation rate in clients can help in designing tests and potentially consolidating the library's APIs.
Knowing that clients on average use a large number of APIs from the library, such as for \sqlite, can help the developers of the library construct tests that encompass many of the library's APIs. 
The data on the API use distribution could be used to focus testing efforts on highly used APIs, while feature requests in little used APIs could be deprioritised.
This information would also allow developers to retire unused APIs and assess the impact of changing an API, as changes to highly used APIs can potentially break many clients. 
As we will see in \S\ref{sec:coverage}, development effort often does not reflect how the library is used in practice.

\vspace{0.06in}


\begin{tcolorbox}[colback=gray!20, colframe=black, boxrule=0.5mm]
  As a library developer, knowing that the average API utilisation is high could indicate that the APIs are tightly coupled together.
  This data can help in designing tests and potentially consolidating the library's APIs.
  The distribution of API uses, including data on the most used and completely unused APIs, can be leveraged to retire APIs, assess the impact of API changes, and prioritise development and maintenance effort.
\end{tcolorbox}



\subsection{API Coverage Analysis}
\label{sec:coverage}

In this part, we aim to understand how API coverage correlates with API implementation size and client usage, and whether clients could be used to improve a library's test suite. 
To answer these questions, we had to restrict ourselves to libraries that include an automated test suite that is easily runnable from the repository of the cloned project and for which we can gather reliable API coverage statistics.
This is typically the test suite used by developers who contribute to the library.
In our set of libraries, we identified cases such as \luajit~\cite{luajit-site} and \freetype~\cite{freetype-site} which do not include an automated test suite in the library repository.  
In \luajit, contributors rely on other test suites to test their pull requests~\cite{luajit-github}.
For \freetype, developers rely on \textit{OSS-Fuzz}~\cite{oss-fuzz} for testing (we confirmed the lack of a test suite by asking the developers~\cite{freetype-testing}).
\ncurses and \hidapi have interactive test suites that require user input. 
In the end, we identified \totalcoverageLibs libraries for which we could easily run their test suites and obtain coverage.

\begin{figure}[tb]
  \centering
  \includegraphics[width=0.5\textwidth]{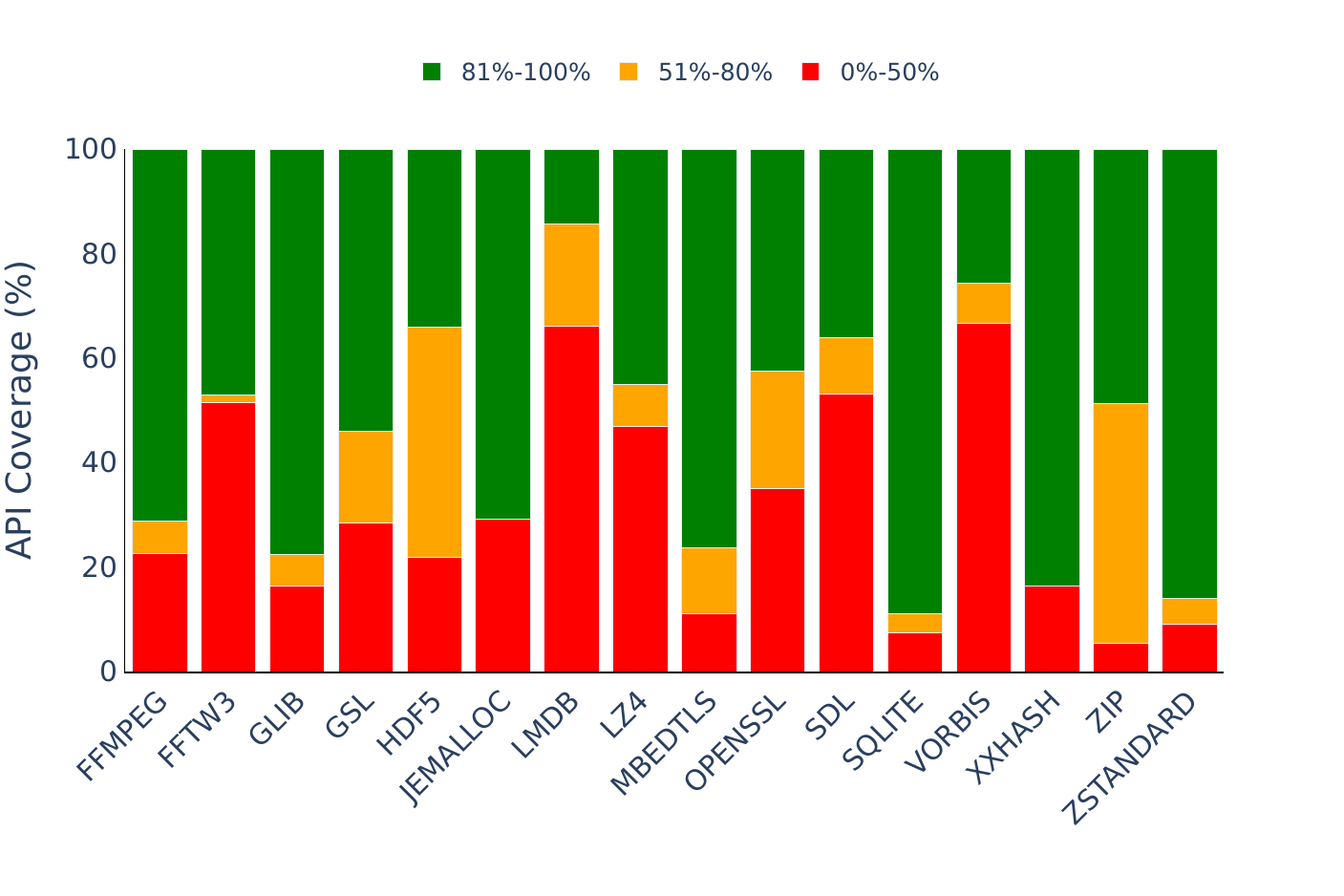}
  \caption{Number of APIs with coverage under 50\%, between 50\% and 80\%, and over 80\%.}
  \label{fig:api_cov_buckets}
\end{figure}

\begin{table}[t]
  \centering
  \caption{API coverage bucketed by API size.
    TCov is the total line coverage in the library code.
    Cov. is the combined line coverage for the API implementations in each size bucket. }
    \begin{tabular}{l|r|r r|r r}
        \toprule
        \multirow{2}{*}{Library} & \multirow{2}{*}{TCov} & \multicolumn{2}{c|}{$\leq 20$} & \multicolumn{2}{c}{$> 20$}\\
        & &  \multicolumn{1}{c}{APIs} & \multicolumn{1}{c|}{Cov.} & \multicolumn{1}{c}{APIs} & \multicolumn{1}{c}{Cov.} \\
        \midrule
        \ffmpeg & \ffmpegcov & \ffmpegapislessthantwenty & \ffmpegapislessthantwentycov & \ffmpegapismorethantwenty & \ffmpegapismorethantwentycov \\
        \fftw & \fftwcov & \fftwapislessthantwenty & \fftwapislessthantwentycov & \fftwapismorethantwenty & \fftwapismorethantwentycov \\
        \glib & \glibcov & \glibapislessthantwenty & \glibapislessthantwentycov & \glibapismorethantwenty & \glibapismorethantwentycov \\
        \gsl & \gslcov & \gslapislessthantwenty & \gslapislessthantwentycov & \gslapismorethantwenty & \gslapismorethantwentycov \\
        \hdf & \hdfcov &  \hdfapislessthantwenty & \hdfapislessthantwentycov & \hdfapismorethantwenty & \hdfapismorethantwentycov \\
        \jemalloc & \jemalloccov & \jemallocapislessthantwenty & \jemallocapislessthantwentycov & \jemallocapismorethantwenty & \jemallocapismorethantwentycov \\
        \lmdb & \lmdbcov  & \lmdbapislessthantwenty & \lmdbapislessthantwentycov & \lmdbapismorethantwenty & \lmdbapismorethantwentycov \\
        \lz & \lzcov  & \lzapislessthantwenty & \lzapislessthantwentycov & \lzapismorethantwenty & \lzapismorethantwentycov \\
        \mbedtls & \mbedtlscov & \mbedtlsapislessthantwenty & \mbedtlsapislessthantwentycov & \mbedtlsapismorethantwenty & \mbedtlsapismorethantwentycov \\
        \ssl & \sslcov & \sslapislessthantwenty & \sslapislessthantwentycov & \sslapismorethantwenty & \sslapismorethantwentycov \\
        \sdl & \sdlcov & \sdlapislessthantwenty & \sdlapislessthantwentycov & \sdlapismorethantwenty & \sdlapismorethantwentycov \\
        \sqlite & \sqlitecov & \sqliteapislessthantwenty & \sqliteapislessthantwentycov & \sqliteapismorethantwenty & \sqliteapismorethantwentycov \\
        \vorbis & \vorbiscov &  \vorbisapislessthantwenty & \vorbisapislessthantwentycov & \vorbisapismorethantwenty & \vorbisapismorethantwentycov \\
        \xxhash & \xxhashcov & \xxhashapislessthantwenty & \xxhashapislessthantwentycov & \xxhashapismorethantwenty & \xxhashapismorethantwentycov \\
        \zip & \zipcov & \zipapislessthantwenty & \zipapislessthantwentycov & \zipapismorethantwenty & \zipapismorethantwentycov \\
        \zstd & \zstdcov & \zstdapislessthantwenty & \zstdapislessthantwentycov & \zstdapismorethantwenty & \zstdapismorethantwentycov \\ 
        \midrule
        \multicolumn{2}{l|}{\textbf{Fully Covered}} & \fullcovlessthantwentyapis &  \fullcovlessthantwentyapispercent & \fullcovmorethantwentyapis & \fullcovmorethantwentyapispercent \\
        \bottomrule
    \end{tabular}%
  \label{tbl:api_coverage}
\end{table}

\mypara{\rqThree}
We run the test suite of each library and compute the coverage achieved in each API implementation.
For \fftw, the developer test suite uses random values for testing, which results in different coverage on each execution.
We ran the test suite nine times and found out that the overall coverage ranges from \fftwcovmin to \fftwcovmax.
The coverage measurements were \fftwcovA, \fftwcovB, \fftwcovC, \fftwcovD, \fftwcovE, \fftwcovF, \fftwcovG, \fftwcovH and \fftwcovI. 
Since the test suite of \fftw is non-deterministic, we considered the median run in graphs and tables.

Figure~\ref{fig:api_cov_buckets} shows how many APIs have coverage under 50\%, between 50\% and 80\%, and over 80\% for each library.
As we can see, there are libraries with generally high API coverage (such as \sqlite and \zstd) and libraries with generally low API coverage (such as \lmdb and \vorbis).
Overall, the number of poorly tested API is high, showing the limited budget available for testing. 

We next contrast API implementation size with API test coverage to understand whether there is any correlation between the two.
Table~\ref{tbl:api_coverage} shows the coverage of the APIs split by size. 
We grouped APIs into two buckets: small APIs with up to 20 ELOC, and larger APIs with over 20 ELOC.
For example, \sdl has \sdlapislessthantwenty APIs that each have up to 20 ELOC with a total coverage, across those APIs, of \sdlapislessthantwentycov; 
and \sdlapismorethantwenty APIs which have each over 20 ELOC, with a total coverage of \sdlapismorethantwentycov.

 

What is evident from Figure~\ref{fig:api_cov_buckets} and Table~\ref{tbl:api_coverage} is that many libraries do not focus on API coverage, although ideally this should be close to 100\% as discussed in \S\ref{sec:intro}.
We also counted the number of fully covered APIs in each group as shown in Table~\ref{tbl:api_coverage}:
\fullcovlessthantwentyapispercent of the \totalapislessthantwenty APIs with up to 20 ELOC have 100\% coverage while only \fullcovmorethantwentyapispercent are fully covered in the greater than 20 ELOC bucket. 

\vspace{0.06in}

\begin{tcolorbox}[colback=gray!20, colframe=black, boxrule=0.5mm]
  Our analysis suggests that many APIs are poorly tested---or not tested at all---and that smaller APIs are easier to test.
  Library developers should use coverage data to direct their testing efforts.
\end{tcolorbox}

\begin{table}[tb]
  \centering
  \caption{APIs used by clients but not tested, sorted in descending order by their percentage in each library.}
  \begin{tabular}{l r r}
    \toprule
    Library & Number of APIs & Percentage  \\
    \midrule
    \vorbis & \vorbisusedanduncoveredapis & \vorbisusedanduncoveredapispercent\\
    \lmdb & \lmdbusedanduncoveredapis & \lmdbusedanduncoveredapispercent\\
    \lz & \lzusedanduncoveredapis & \lzusedanduncoveredapispercent\\
    \ssl & \sslusedanduncoveredapis & \sslusedanduncoveredapispercent\\
    \sdl & \sdlusedanduncoveredapis & \sdlusedanduncoveredapispercent\\
    \fftw & \fftwuncovered & \fftwusedanduncoveredapispercent\\
    \jemalloc & \jemallocusedanduncoveredapis & \jemallocusedanduncoveredapispercent\\
    \xxhash & \xxhashusedanduncoveredapis & \xxhashusedanduncoveredapispercent\\
    \ffmpeg & \ffmpegusedanduncoveredapis & \ffmpegusedanduncoveredapispercent\\
    \mbedtls & \mbedtlsusedanduncoveredapis & \mbedtlsusedanduncoveredapispercent\\
    \zstd & \zstdusedanduncoveredapis & \zstdusedanduncoveredapispercent\\
    \sqlite & \sqliteusedanduncoveredapis & \sqliteusedanduncoveredapispercent\\
    \glib & \glibusedanduncoveredapis & \glibusedanduncoveredapispercent\\
    \gsl & \gslusedanduncoveredapis & \gslusedanduncoveredapispercent\\
    \hdf & \hdfusedanduncoveredapis & \hdfusedanduncoveredapispercent\\
    \zip & \zipusedanduncoveredapis & \zipusedanduncoveredapispercent\\
    \bottomrule
  \end{tabular}
  \label{tbl:uncoveredapis}
\end{table}
\begin{figure*}[t]
  \scalebox{0.9}{ 
  \begin{minipage}{\textwidth}
  \centering
  \captionsetup[subfigure]{skip=0pt}
  \begin{minipage}{\textwidth} 
    \hspace{1cm} 
    \begin{subfigure}[t]{0.23\textwidth}
        \captionsetup{labelformat=empty, justification=centering, singlelinecheck=off}
        \includegraphics[width=\linewidth]{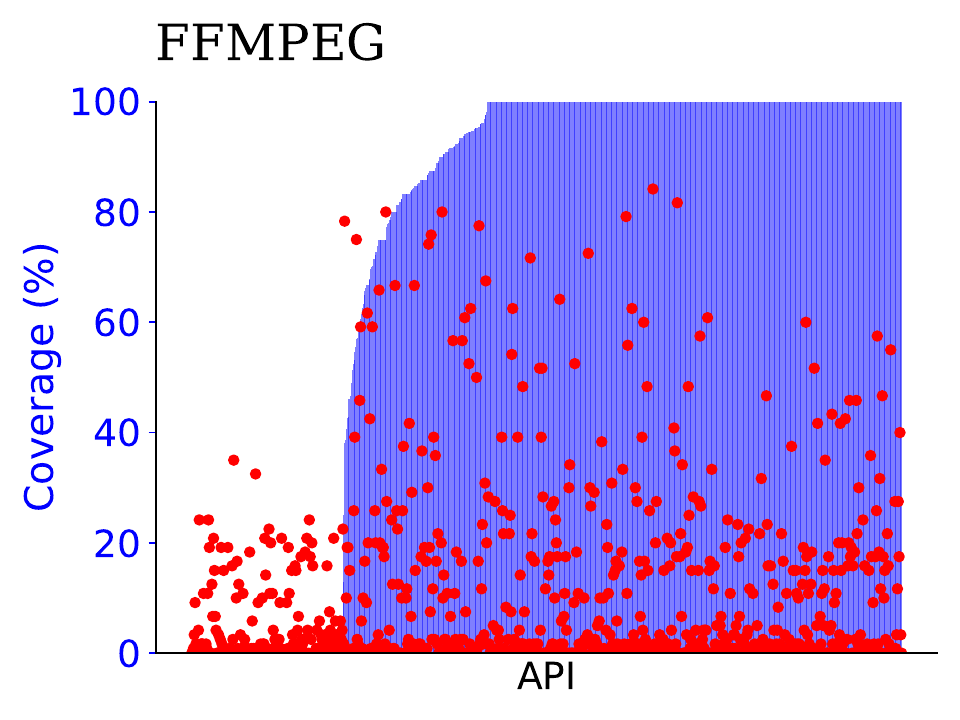}
        \label{subfig:ffmpeg_coverage}
    \end{subfigure}
    \begin{subfigure}[t]{0.23\textwidth}
        \captionsetup{labelformat=empty, justification=centering}
        \includegraphics[width=\linewidth]{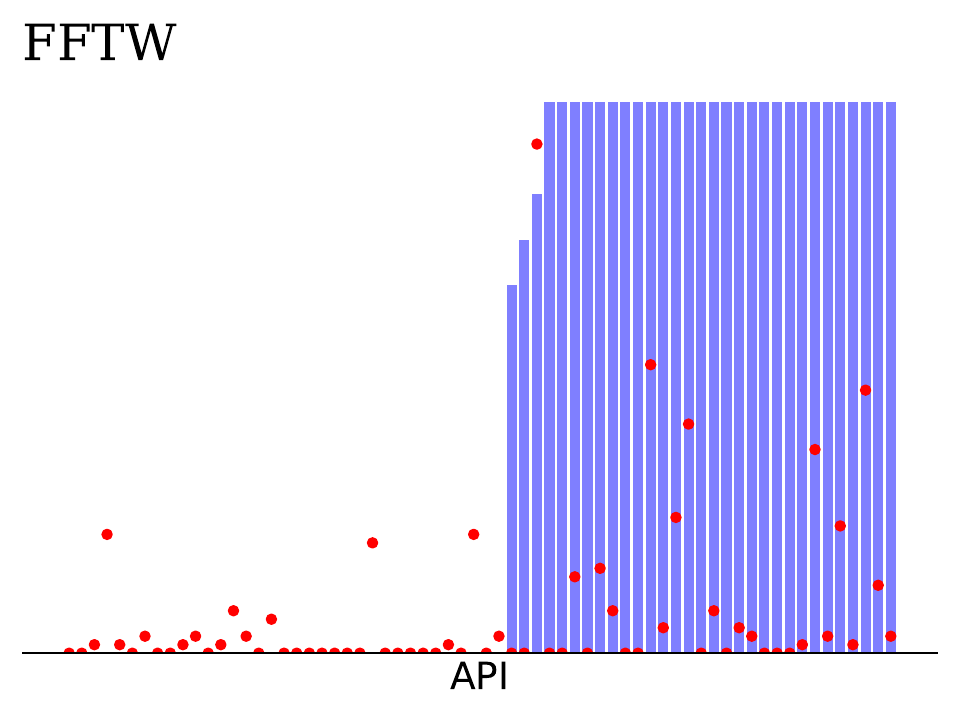}
        \label{subfig:fftw3_coverage}
    \end{subfigure}
    \begin{subfigure}[t]{0.23\textwidth}
        \captionsetup{labelformat=empty, justification=centering}
        \includegraphics[width=\linewidth]{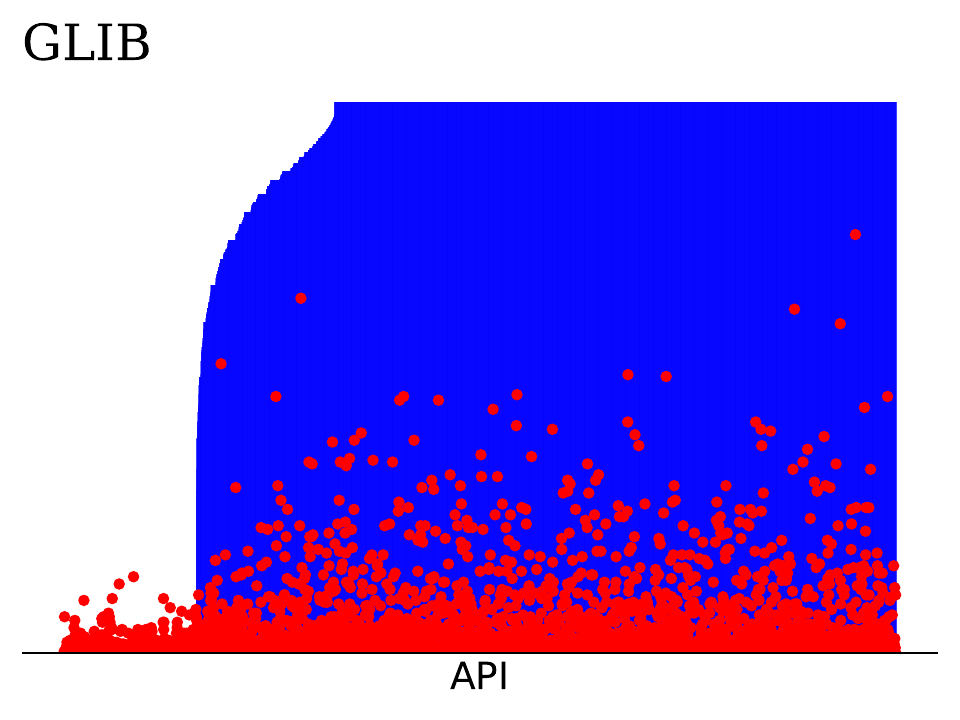}
        \label{subfig:glib_coverage}
    \end{subfigure}
    \begin{subfigure}[t]{0.23\textwidth}
      \captionsetup{labelformat=empty, justification=centering}
      \includegraphics[width=\linewidth]{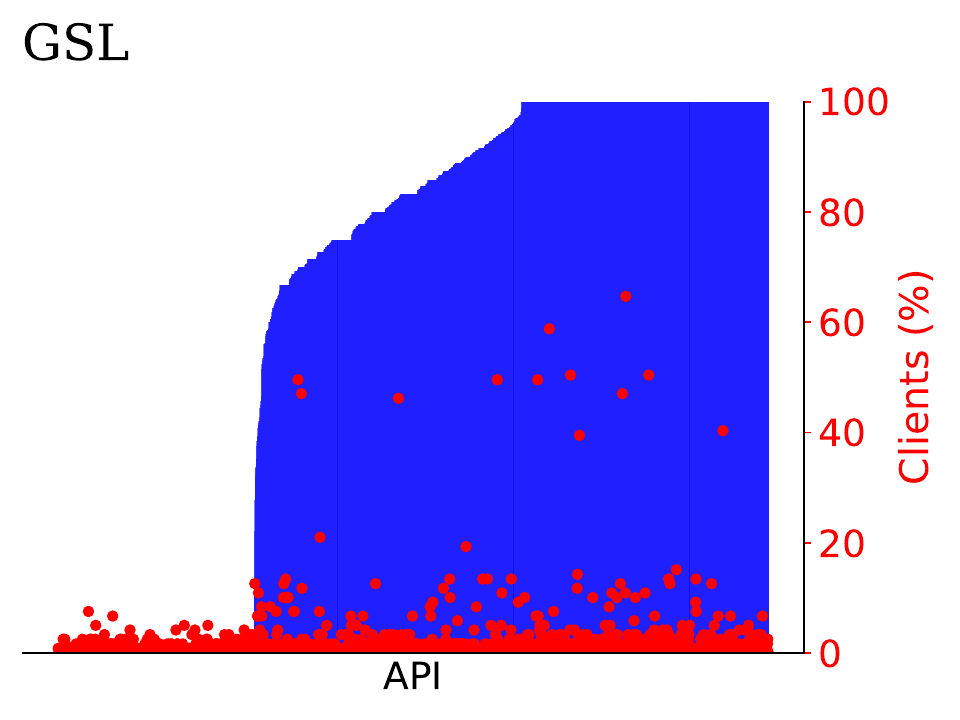}
      \label{subfig:gsl_coverage}
    \end{subfigure}   
  \end{minipage}

  \vspace{-0.5cm} 

  \begin{minipage}{\textwidth} 
    \hspace{1cm} 
    \begin{subfigure}[t]{0.23\textwidth}
      \captionsetup{labelformat=empty, justification=centering}
      \includegraphics[width=\linewidth]{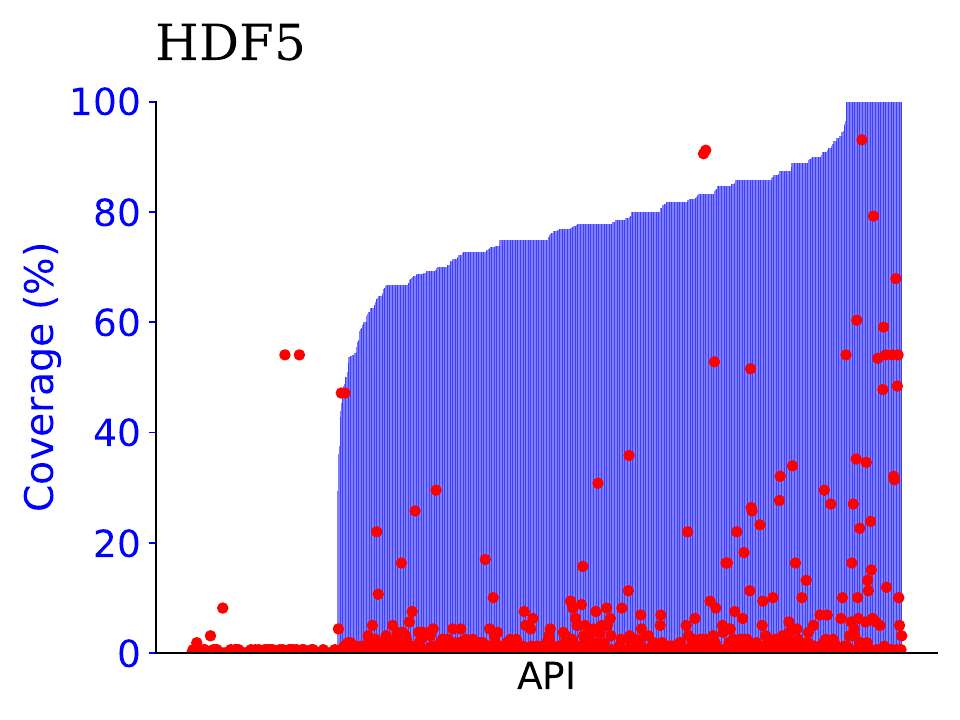}
      \label{subfig:hdf_coverage}
    \end{subfigure}
    \begin{subfigure}[t]{0.23\textwidth}
        \captionsetup{labelformat=empty, justification=centering}
        \includegraphics[width=\linewidth]{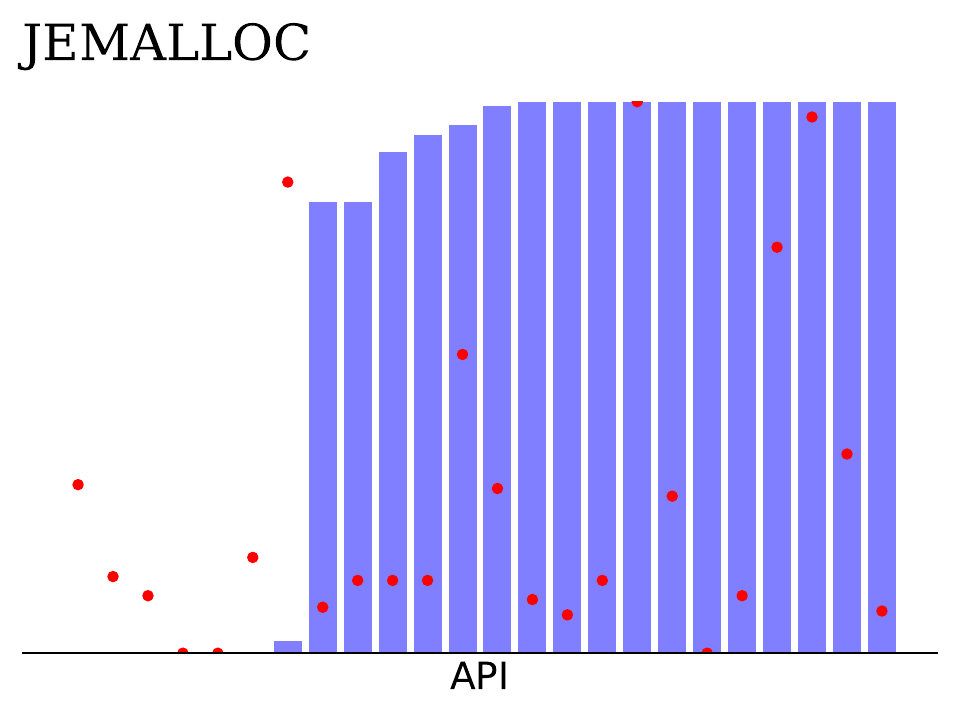}
        \label{subfig:jemalloc_coverage}
    \end{subfigure}
    \begin{subfigure}[t]{0.23\textwidth}
      \captionsetup{labelformat=empty, justification=centering}
      \includegraphics[width=\linewidth]{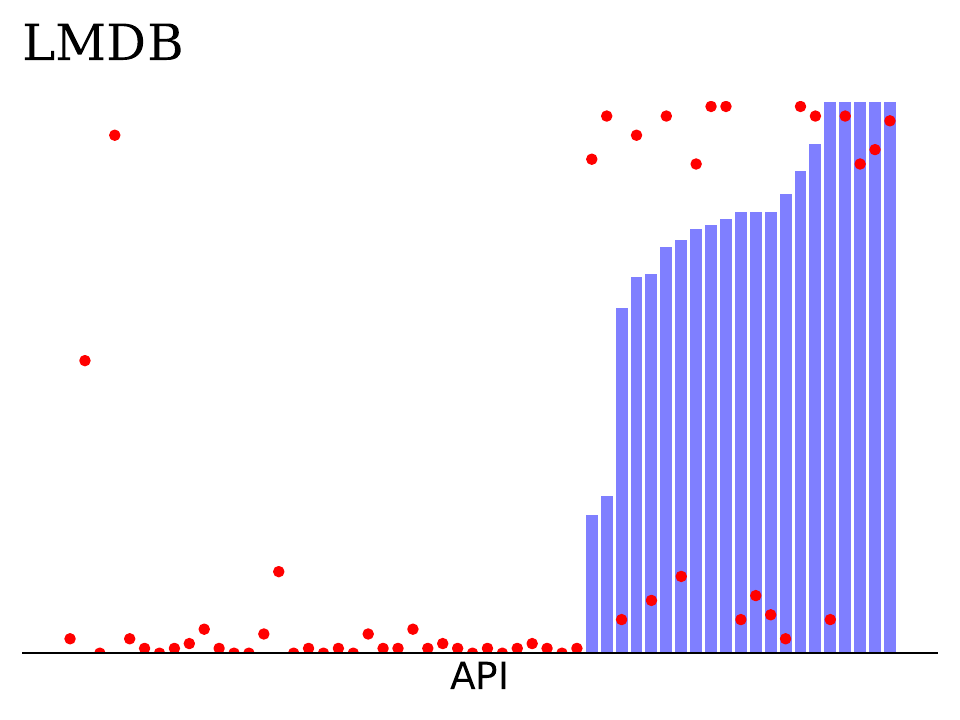}
      \label{subfig:lmdb_coverage}
    \end{subfigure}
    \begin{subfigure}[t]{0.23\textwidth}
      \captionsetup{labelformat=empty, justification=centering}
      \includegraphics[width=\linewidth]{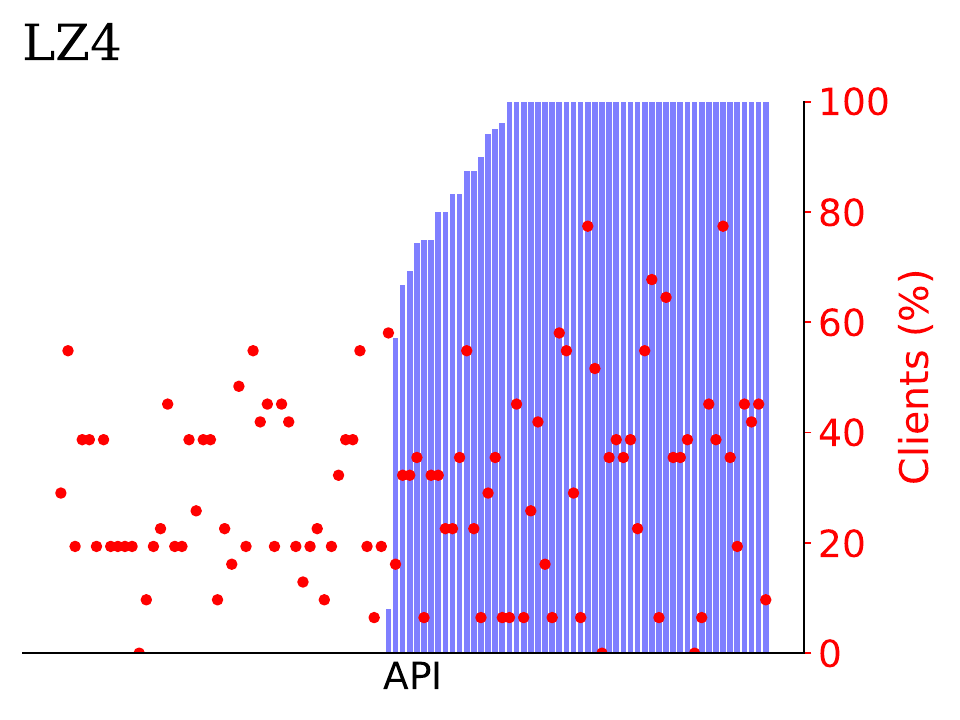}
      \label{subfig:lz4_coverage}
    \end{subfigure}
  \end{minipage}
  
  \vspace{-0.5cm} 

  \begin{minipage}{\textwidth} 
    \hspace{1cm} 
    \begin{subfigure}[t]{0.23\textwidth}
      \captionsetup{labelformat=empty, justification=centering}
      \includegraphics[width=\linewidth]{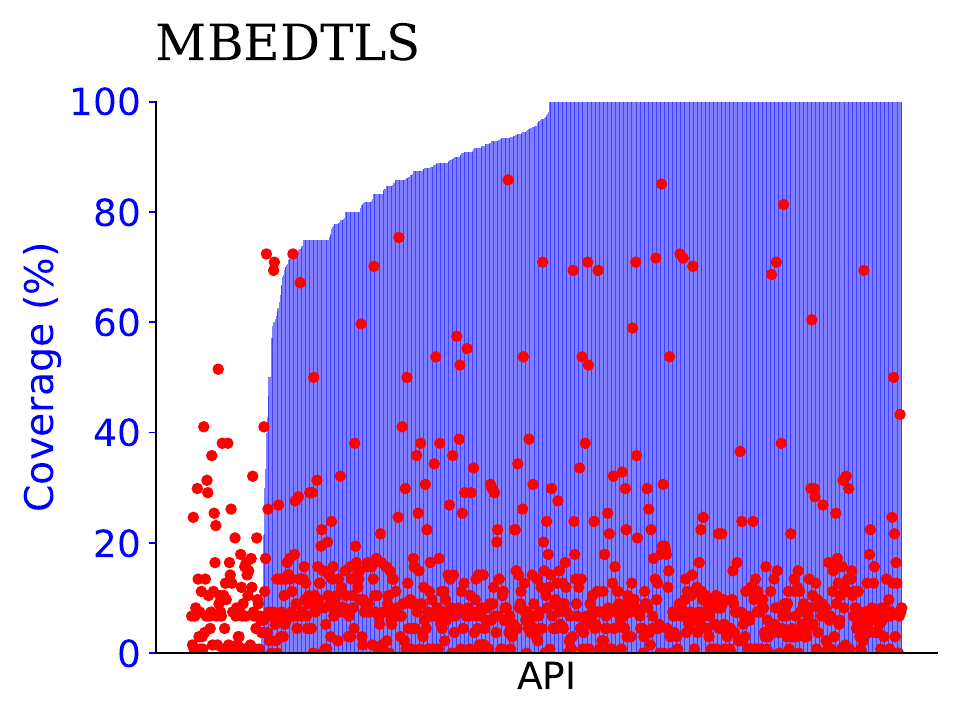}
      \label{subfig:mbedtls_coverage}
    \end{subfigure}
    \begin{subfigure}[t]{0.23\textwidth}
      \captionsetup{labelformat=empty, justification=centering}
      \includegraphics[width=\linewidth]{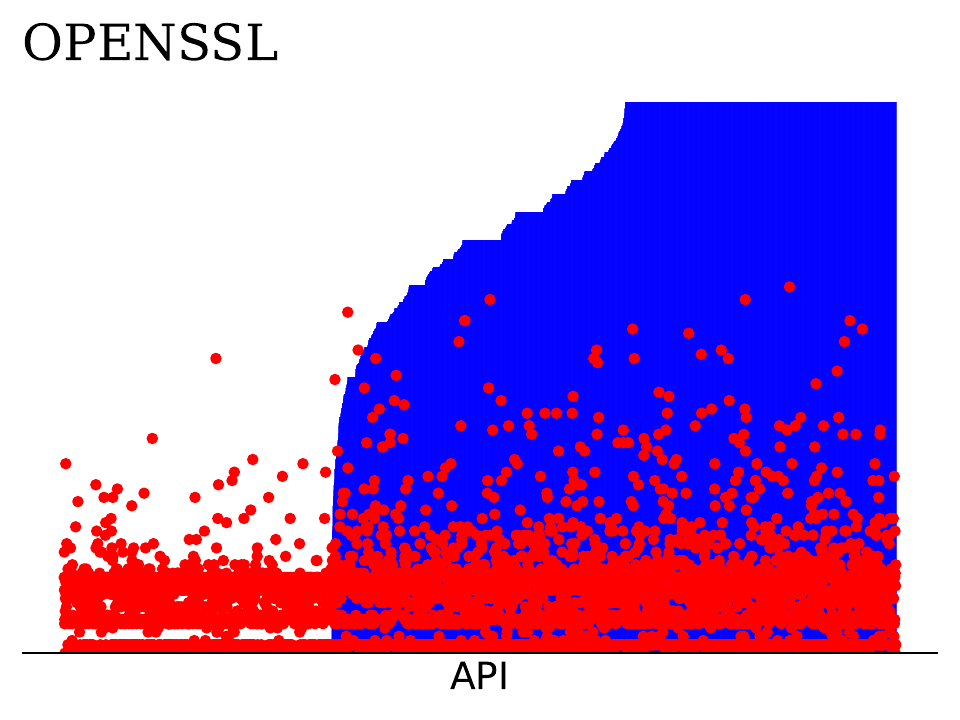}
      \label{subfig:ssl_coverage}
    \end{subfigure}
    \begin{subfigure}[t]{0.23\textwidth}
        \captionsetup{labelformat=empty, justification=centering}
        \includegraphics[width=\linewidth]{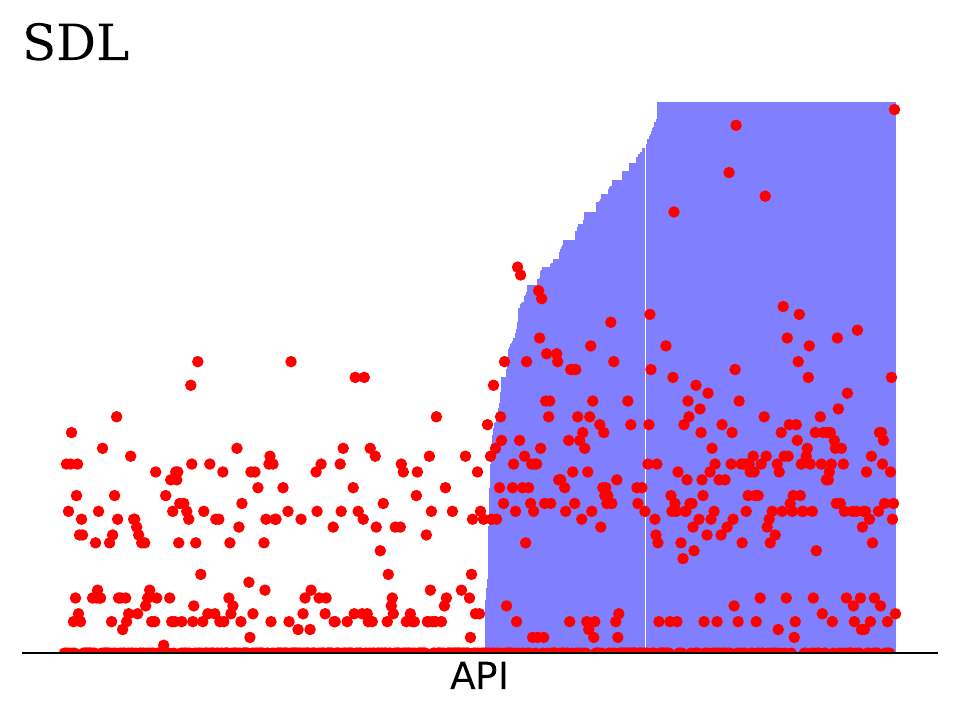}
        \label{subfig:sdl_coverage}
    \end{subfigure}
    \begin{subfigure}[t]{0.23\textwidth}
        \captionsetup{labelformat=empty, justification=centering}
        \includegraphics[width=\linewidth]{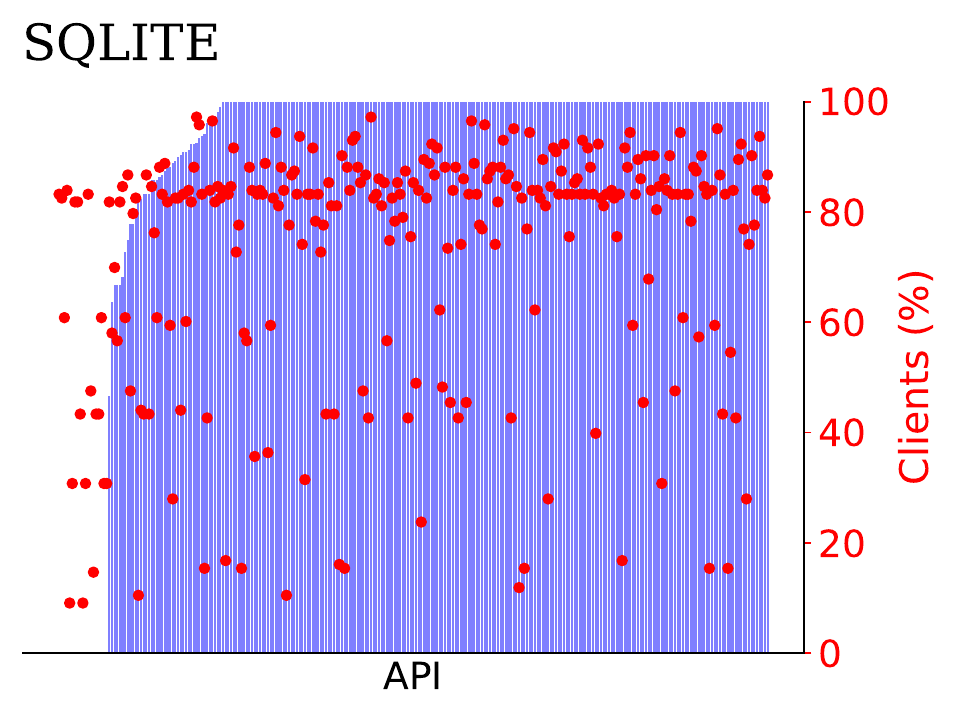}
        \label{subfig:sqlite_coverage}
    \end{subfigure}   
  \end{minipage}

  \vspace{-0.5cm} 

  \begin{minipage}{\textwidth} 
    \hspace{1cm} 
    \begin{subfigure}[t]{0.23\textwidth}
        \captionsetup{labelformat=empty, justification=centering}
        \includegraphics[width=\linewidth]{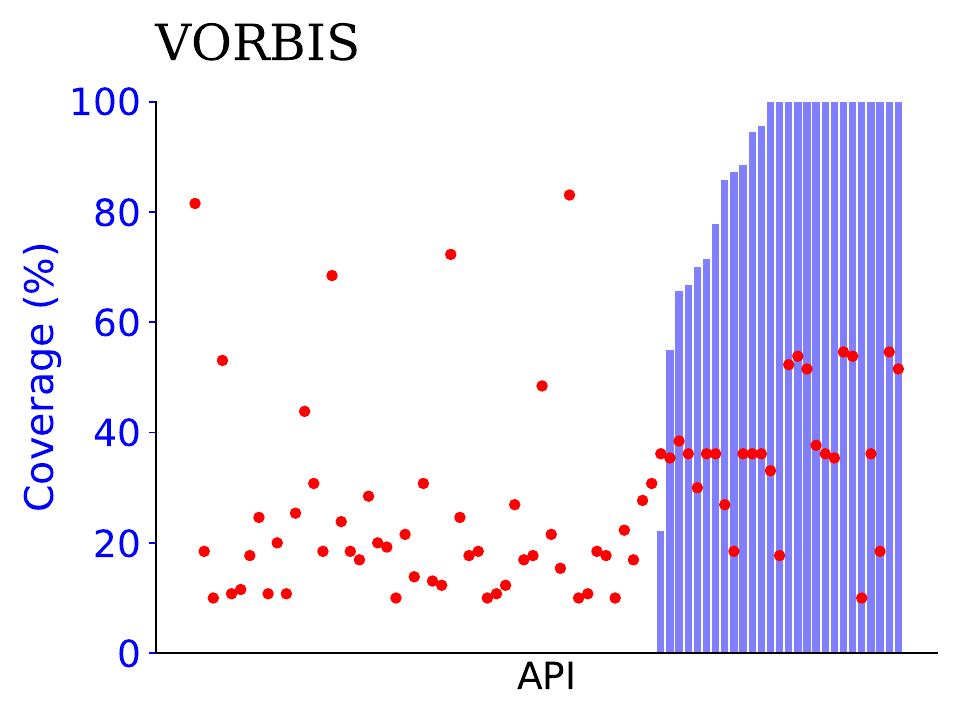}
        \label{subfig:vorbis_coverage}
    \end{subfigure}
    \begin{subfigure}[t]{0.23\textwidth}
        \captionsetup{labelformat=empty, justification=centering}
        \includegraphics[width=\linewidth]{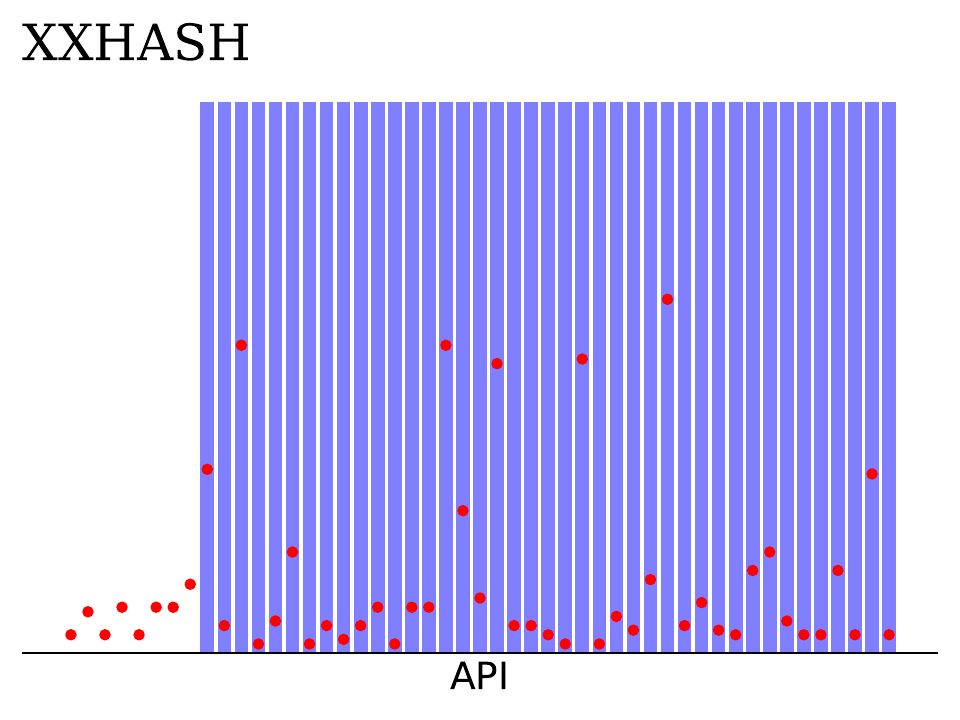}
        \label{subfig:xxhash_coverage}
    \end{subfigure}
    \begin{subfigure}[t]{0.23\textwidth}
        \captionsetup{labelformat=empty, justification=centering}
        \includegraphics[width=\linewidth]{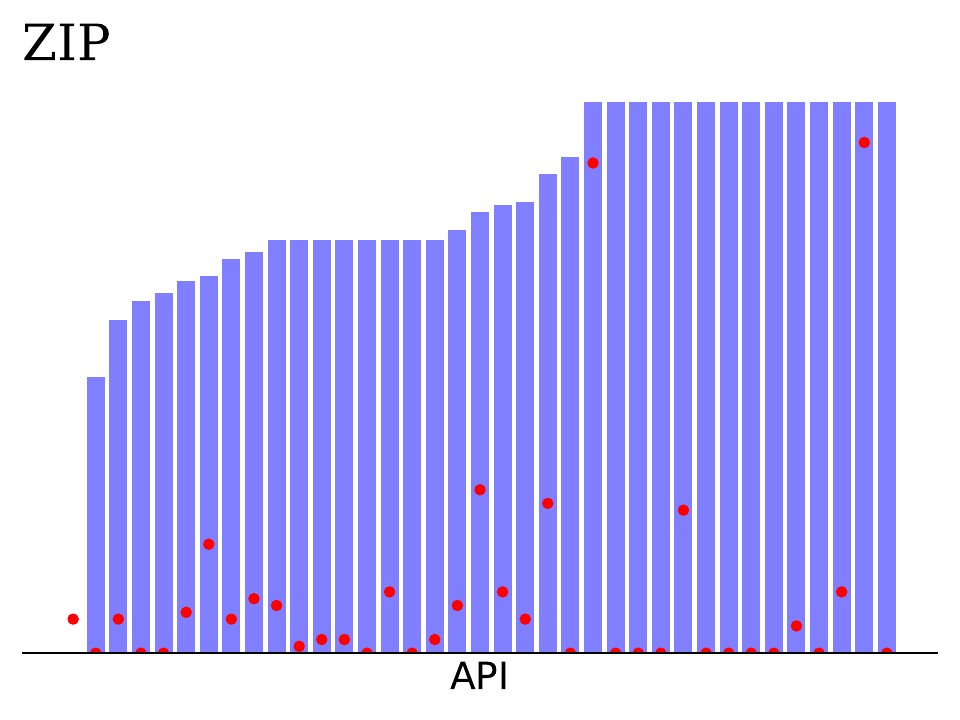}
        \label{subfig:zip_coverage}
    \end{subfigure}
    \begin{subfigure}[t]{0.23\textwidth}
      \captionsetup{labelformat=empty, justification=centering}
      \includegraphics[width=\linewidth]{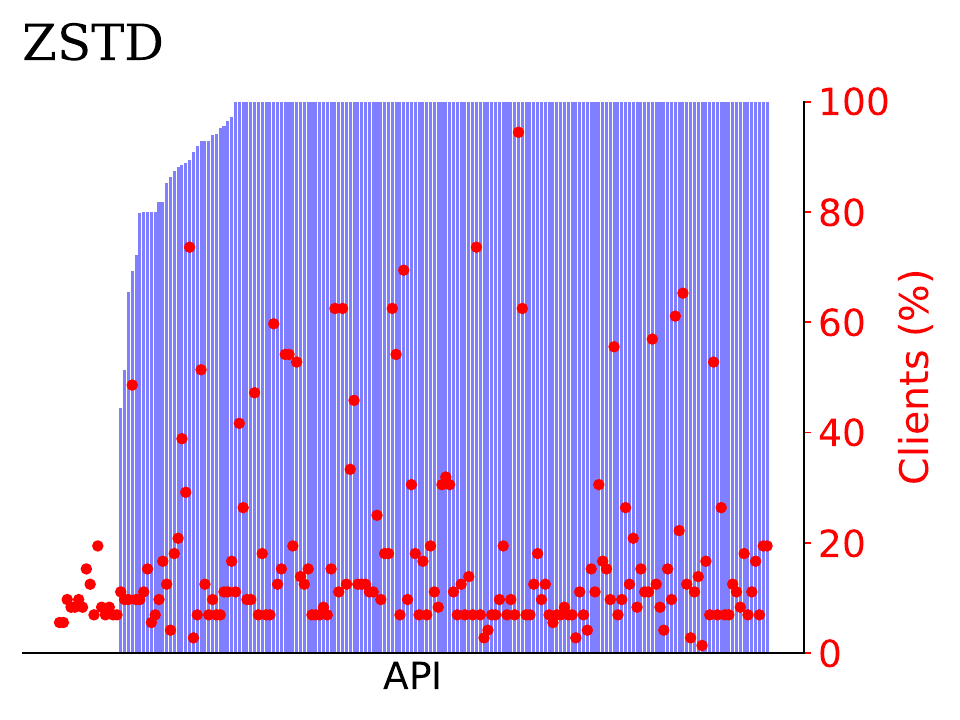}
      \label{subfig:zstd_coverage}
    \end{subfigure}   
  \end{minipage}
\end{minipage}
  }
  \caption{API coverage for each library. The blue bars represent the coverage for each API, while the red dots show the percentage of clients using that API. }
  \label{fig:api_coverage}
\end{figure*}

\mypara{\rqFour}
An important question is whether APIs which are widely used in the field are also well-tested.
Ideally, developers would spend more resources writing tests for popular APIs, and fewer for those APIs which are rarely used.

Figure~\ref{fig:api_coverage} shows coverage of each API against usage in our data set. 
The blue bars show the coverage of each \apisize, with the bars sorted in ascending order.
For each API, the figure also shows, as a red dot, the percentage of clients that use that API.
In almost all libraries, we can see a discrepancy between tested APIs and usage by clients. 
Very few library, such as \zstd and \zip,  have good overall testing of their APIs relative to usage. 
For example, \lmdbapiwithclientsnocov, from \lmdb, is used by \lmdbapiwithclientsnocovpercent of the library's clients yet it is not tested by the library's test suite. 
Similarly, \vorbisapiwithclientsnocov, from \vorbis, is used by \vorbisapiwithclientsnocovpercent of the library's clients, with no test coverage.

Table~\ref{tbl:uncoveredapis} shows for each library the number of APIs used but not tested by the library test suite, and their percentage from the total number of library APIs.
As can be seen, these percentages vary from only 3\% in \zip to 65\% in \vorbis, with a median of 15\%.

In Figure~\ref{fig:api_coverage} we can also see cases where the number of clients is low yet the coverage is high.
While this is not a problem per se, the overall picture shows that developers could do a better job prioritising their testing efforts if they had information about how their APIs are used in the field.

\vspace{0.06in}

\begin{tcolorbox}[colback=gray!20, colframe=black, boxrule=0.5mm]
  There is a clear discrepancy between API usage by clients and \apisize test coverage. 
  Library developers should use API usage and coverage information to better prioritise their efforts and ensure the most used APIs are well tested and optimised. 
\end{tcolorbox}

\begin{table}[tb]
  \centering
  \caption{API coverage improvement achieved by the clients' test suites.
    For each library, we list the total extra coverage added to the library, the number of newly covered APIs, and the number of APIs where coverage was improved.}
  \begin{tabular}{ll|r|rr}
    \toprule
    \multirow{2}{*}{Library} & \multirow{2}{*}{Client} & Extra     & \multicolumn{2}{c}{APIs}  \\
                                         &                                     & TCov & New  & Improved \\
    \midrule
    \lmdb & \knot & \lmdbimprovedcov & \lmdbnewlycoveredapis & \lmdbimprovedcovapis \\
    \vorbis & \sfml & \vorbisimprovedcov & \vorbisnewlycoveredapis & \vorbisimprovedcovpis \\
    \fftw & \cava & \fftwimprovedcov & \fftwnewlycoveredapis & \fftwimprovedcovapis \\
    \sdl & \ufo & \sdlimprovedcov & \sdlnewlycoveredapis & \sdlimprovedcovapis \\
    \bottomrule
  \end{tabular}%
  \label{tbl:improvedcov}
\end{table}

\mypara{\rqFive}
RQ2 and RQ3 have shown that many APIs used by clients are poorly tested, or not tested at all, by the library test suite.
We show that many clients already incorporate tests that exercise the library APIs, and those tests could be used by library developers to enhance testing of the library.
In previous work, we explored techniques for extracting these tests automatically from client codebases, such that they can be used in the library's test suite without any dependence on the client code or other libraries~\cite{apislicer}.

Since we are interested in evaluating whether we can improve API coverage in libraries, we focused our attention on libraries that would benefit the most from this; libraries that have more than 20\% of their APIs used by clients but not tested by library test suites.
Looking at Table~\ref{tbl:uncoveredapis} we selected six libraries for our evaluation; \vorbis, \lmdb, \lz, \ssl, \sdl, and \fftw. 
For each of these libraries, we identify the top ten clients and use them to check if running their test suites would lead to an increase in API coverage.
We exclude any clients that are complex to build, such as \textit{FreeBSD}, could not be built on a Linux system, or do not have a test suite.
For each client we attempt to build and run the test suite to see if there is an increase in API coverage in the target library.
Once we find a client that increases API coverage in the target library we report it and stop. 

For \lz and \ssl we were not able to improve coverage by building and running the test suites of the top 10 clients. 
It is worth noting that \lz had one of the lowest number of actual clients using the library.
Several of those clients were operating systems or required specific hardware (such as recent GPUs), which made it difficult to build them.

We were able to increase API coverage in \lmdb, \vorbis, \fftw and \sdl.
For \lmdb we used \knot~\cite{knot-site}, a high-performance authoritative-only DNS server; 
for \vorbis we used \sfml (Simple and Fast MultiMedia Library)~\cite{sfml-site};
for \fftw we used \cava~\cite{cava-site}, a cross-platform audio visualizer; and for \sdl we used a game called \ufo~\cite{ufo-site}.

Table~\ref{tbl:improvedcov} shows the improvements in terms of coverage.
For each library, we list the number of APIs which were \textit{newly covered} as a result of executing the client test suite and the number of APIs where improved coverage was achieved.


For \sdl, we cover six previously uncovered APIs: two become fully covered (100\%) while the other four achieve coverage of over 60\%.
For \lmdb, we improve the testing of eight APIs, four of which were previously uncovered.
Two of the eight APIs become fully covered, one of them uncovered before.
For \vorbis, we improve testing of fifteen APIs.
Six of them become fully covered, all of them uncovered before.
For \fftw, which has a non-deterministic test suite, we ran \cava's test suite nine times and consistently cover two new APIs that were previously uncovered: \fftwapione and \fftwapitwo.

For \lmdb and \vorbis, \lmdbnewlycoveredlinesapis and \vorbisnewlycoveredlinesapis new lines were covered in the API implementations of the libraries respectively. 
API line coverage in \vorbis went from 35.9\% to 51.3\% while in \lmdb from 26.6\% to 44.0\%.
For the other two benchmarks, the numbers of extra new lines covered was less substantial.

\vspace{0.06in}

\begin{tcolorbox}[colback=gray!20, colframe=black, boxrule=0.5mm]
  In summary, we were able improve API coverage using client test suites in 4 out of 6 libraries. 
This shows that library developers can leverage clients that use their libraries to generate tests that reflect how the library APIs are used in practice. 
\end{tcolorbox}

\subsection{Threats to Validity}
\label{sec:threats}
Our study has several threats to validity.


To handle the large number of clients, we used textual search for API use identification (see \S\ref{sec:usages}), which may introduce some false positives and negatives.
We tried to minimise these issues by conducting an initial study to choose the most precise scalable method.
Based on our observations, we believe such instances are not significant enough to invalidate the insights gathered from our empirical study.

When we measure size and line coverage for the APIs, we restrict our measurement to the lines in the implementation of the API entry functions---see \S\ref{sec:lib-proc} for an extended discussion.

The conclusions of our study might not generalise beyond the libraries and clients examined.
However, our \totalLibs libraries are representative of popular C libraries and the clients considered per library are high enough in number and popularity to capture how the libraries are used in practice. 


As with any large-scale study, we cannot dismiss the possibility of implementation errors. 
To mitigate this threat, we double-checked the numbers, and provide an artifact.

\section{Related Work}
\label{sec:related}
To our knowledge, our study is the first that contrasts usage and test coverage of C libraries in the C/C++ ecosystem.
We study \totalLibs libraries with the objective of offering valuable insights to library developers, rather than to the users of the libraries.


There are several studies that look at API usage in the \java ecosystem.
Qiu \etal~\cite{japiextractor} performed an empirical study on API usage for 5,000 open-source \java projects.
This study looked at usage across core and third-party \java libraries.
It analysed 16,329 distinct third-party libraries and found that only 15 were used by over 10\% of the projects in their data set, while 265 were used by 1\%, and 9,830 by a single project each.
In our study, out of 2,520 dependencies, 82\% had less than 10 users, while 14\% had 10 to 49 users, and 2\% had 50 to 99 users. 
As we showed in Figure~\ref{fig:usagedistrib}, only two dependencies had more than 1,000 users. 
The study also investigated the usage of the \java core library by measuring how much of the library is used by projects in the corpus. 
The paper reports that 41.2\% of the core methods of \java8 were never used across their client corpus. 
In our analysis, out of the \totalLibs libraries, four had 50\% or more of their API unused.
The study was conducted with respect to packages and classes, which is not directly comparable to API usage in C libraries. 

Other studies analyse library usage from the \textit{clients' perspective}~\cite{uppdatera,api-usages-empirical-study}.
Hejderup and Gousios~\cite{uppdatera} assess the effectiveness of \java project test suites in  covering usages of third-party libraries.
The aim of the work is to assess how reliable test suites are as a means to evaluate the compatibility of updated library versions.
The study did not explore if client test suites can be leveraged to improve library coverage, instead it focused on determining whether clients test all the APIs they use from direct dependencies.

Zhong and Mei~\cite{api-usages-empirical-study} investigate how seven \java applications use internal and external APIs.
The emphasis of the study is on understanding how clients use APIs by understanding common ways clients call different types of elements of an API.
The paper studies how frequently APIs are used and find out that for most libraries only a small portion of APIs are called.
We perform our study on \actualNonUniqueClients C/C++ clients and show in \S\ref{subsec:client_usage_analysis} that for most libraries average client usage sits below 40\%. 
But we also show that full utilisation of the API is possible for some libraries.



Harrand \etal~\cite{harrand2022api} analysed API usage for 94 \java libraries across 829,410 clients. 
The objective of their research was to explore the contradiction between Hyrum's Law and findings that show that for most libraries only a fraction of their APIs are used by clients. 
Our evaluation shows that five libraries had full API usage. 
Our findings align with those of Harrand \etal~\cite{harrand2022api} in that in some cases, with enough clients, all APIs of a library are used (Hyrum's Law), yet at the same time most APIs can be significantly reduced and still fulfil the needs of the majority of the clients.


Schittekat \etal~\cite{schittekat2022can} conducted an evaluation on four Python packages across 14 clients to assess whether using the tests from the clients can improve coverage in the libraries. 
They were able to improve coverage in two out of the four packages.
The improvements range from a maximum of 28\% increase to a minimum of 1\%. 
The paper did not investigate whether new APIs were covered from the Python packages.

Prior work has also analysed library usage with the objective of detecting breaking changes~\cite{mujahid2020using,api-stability-empirical-android,noregrets}.
Mujahid \etal\cite{mujahid2020using} performed an empirical study on 391,553 \textit{npm} packages to evaluate if the tests from client projects can be used to detect breaking changes in packages.
The paper found that client tests can cover up to 47\% of the code for the target dependency but did not look at improving the coverage of the target dependencies, which is what we show is possible in our study.
McDonnell \etal\cite{api-stability-empirical-android} conduct an empirical study to understand how clients using the Android SDK keep up with changes in the API.
The study looks at how APIs provided by the Android SDK change over time and how clients catch up with those changes. 
Similar to \cite{mujahid2020using} and \cite{noregrets}, the study is more concerned with how changes in the library, as it evolves, impacts clients.

\section{Conclusion}
\label{sec:conclusion}
Libraries represent an indispensable component of the software ecosystem.
Unfortunately, library developers often have little knowledge of how their code is used in practice.
In this paper, we present a large-scale empirical study in which we analyse API usage across \totalLibs C libraries and \actualNonUniqueClients C/C++ clients.
We developed \libprobe, a lightweight analysis framework that can provide valuable insights to library developers regarding how their APIs are used in the field, to help them prioritise their efforts in maintaining, improving, and testing their libraries. 
Our study shows that library developers do not prioritise their effort based on how clients use their APIs---popular APIs are often poorly tested, with rarely-used ones well tested instead.
We further show that client test suites can be leveraged to improve library testing, with the important advantage that those tests are representative of how the APIs are used in the field.

\section{Acknowledgements}
We thank Martin Nowack for his feedback on the text.
This research has received funding from from the Engineering and Physical Sciences Research Council (EPSRC) via a PhD studentship, and from the European Research Council (ERC) under the European Union's Horizon 2020 research and innovation programme (grant agreement 819141).

\balance
\bibliographystyle{ACM-Reference-Format}
\bibliography{references,cadar,cadar-crossrefs,cadar-macros}
\end{document}